\documentclass[10pt,aps,pre,onecolumn,nofootinbib,showpacs,longbibliography,superscriptaddress,notitlepage]{revtex4-1}
\usepackage{graphicx}
\usepackage{amssymb,amsmath,amsfonts,wasysym}
\usepackage{wrapfig}
\usepackage{physics}
\usepackage{subfigure}
\usepackage{color}
\usepackage{floatrow}
\usepackage{listings}
\usepackage{color}
\usepackage{epigraph}
\usepackage{yfonts}

\newcommand{\beq}{\begin{equation}}
\newcommand{\eeq}{\end{equation}}
\newcommand{\ba}{\begin{align}}
\newcommand{\ea}{\end{align}}

\begin{document}

\title{From the replica \emph{trick} to the replica symmetry breaking \emph{technique}}
\author{Patrick Charbonneau}
\affiliation{Departments of Chemistry and Physics, Duke University, Durham, North Carolina, 27708, USA}
\email{patrick.charbonneau@duke.edu}
\date{\today}
\begin{abstract}
Among the various remarkable contributions of Giorgio Parisi to physics, his formulation of the replica symmetry breaking solution for the Sherrington-Kirkpatrick model stands out. In this article, different historical sources are used to reconstruct the scientific and professional contexts of this prodigious advance.
\end{abstract}
\date{\today}
\maketitle

\epigraph{A \emph{trick} is a clever idea that can be used once, while a \emph{technique} is a mature trick that can be used at least twice. We will see in this section that tricks tend to evolve naturally into techniques.}{D. Knuth, \emph{The Art of Computer Programming}, Vol. 4A (2014),  Section 7.1.3}

\section{Introduction}
Giorgio Parisi was granted half of the 2021 Nobel Prize in Physics ``for the discovery of the interplay of disorder and fluctuations in physical systems from atomic to planetary scales.'' This unusually vague citation reflects the depth and breadth of Parisi's seminal contributions to physics, which span from quantum chromodynamics to climate change and bird flocking. Yet it partly obfuscates that without one particular such contribution his career might not have stood out as markedly as it did. For some, Parisi's \emph{replica symmetry breaking} (RSB) solution of the Sherrington-Kirkpatrick model of spin glasses and his ensuing work on disordered systems deserve particular note. 
Various other award citations support that impression\footnote{For instance, the 1992 Boltzmann Medal was ``awarded to Giorgio Parisi for his fundamental contributions to statistical physics, and particularly for his solution of the mean field theory of spin glasses,'' and the 2005 Dannie Heineman Prize for Mathematical Physics was ``for fundamental theoretical discoveries in broad areas of elementary particle physics, quantum field theory, and statistical mechanics; especially for work on spin glasses and disordered systems''.}, and even the scientific background for this latest prize is somewhat more explicit~\cite{nobel2021}. A possible explanation for this discrepancy might be that, despite being in its fifth decade, RSB is still not uncontroversial. From a mathematical standpoint, it indeed remains a non-rigorous (and therefore somewhat uncontrolled) calculation scheme. 

In a nutshell, RSB follows from using the replica identity to average the free energy of a system over some \emph{quenched} (or frozen) disorder,
\begin{equation}
\label{eq:replica}
-\beta \overline{F}=\overline{\log Z}=\lim_{n\rightarrow 0} \frac{\partial \overline{Z^n}}{\partial n},
\end{equation}
given the partition function $Z$ for a system with a given realization of that disorder. Why this particular average? An early articulation of the physics of quenched disorder was offered in the late 1950s by theoretical physicist Robert Brout (1928-2011)~\cite{brout1959}, who like many others at the time was considering the impact of randomness in solid state physics. By then, systems with \emph{annealed} impurities were well understood. Such impurities are deemed equilibrated in all instances, and therefore the free energy of such a system can be obtained from first averaging the partition function, i.e., $-\beta F=\log \overline{Z}$. Brout noted:
\begin{quotation}
We are, however, interested not in the free energy of such a system, but rather the free energy of a given system of [impurities] frozen into their positions [\ldots]. Since we are interested in a random sample, it is clear that it is desired first to calculate the spin sum in the partition function (since the spin system is assumed to be in thermal equilibrium) and then to average the logarithm of this quantity over all spatial configurations. In short, we must calculate $\overline{\log Z}$, i.e., the mean free energy.~\cite{brout1959}
\end{quotation}
Even with this solid physical reasoning, Eq.~\eqref{eq:replica} remains problematic. The quantity $Z^n$ can only be expediently computed for an integer number of system replicas $n$, while the analytic continuation to the reals is not guaranteed to be unique. Empirically, when it isn't, new physics emerges. Technically speaking, averaging $Z^n$ can couple the $n$ replicas. If all these couplings (or overlaps $q$) are equivalent the continuation is well defined and the solution is said to be replica symmetric (RS). If this coupling breaks the symmetry between replicas, however, particular care must be taken. The Parisi ansatz considers a subgroup of symmetries for breaking the equivalence between replicas that solves whole classes of models. In many cases that solution has been independently (albeit onerously) demonstrated to be exact (see, e.g., Refs.~\cite{talagrand2010,talagrand2011}); in many others it is presumed correct but the mathematical jury is still out. Notwithstanding its limited rigor, the RSB approach has markedly impacted our understanding of disordered systems in physics and beyond~\cite{farbeyond2023}. 

As part of an ongoing oral history project, Francesco Zamponi and I are trying to understand the intellectual, sociological and pedagogical sources and impacts of this particular tension between efficacy and rigor in theoretical physics~\cite{caphes2022}. The present article does not attempt to summarize the whole project, but to more narrowly capture how the various formulations and uses of the replica trick eventually brought Parisi to the correct RSB technique. 

\section{$n\rightarrow0$ Identity in Mathematics: Hardy and Reisz}
In 1928, at the end of his first two-year term as president of the London Mathematical Society, British mathematician G.~H.~Hardy (1877-1947)
announced his plan to publish a chapter dedicated specifically to inequalities. He found the topic to be ```bright' and amusing, and intelligible without large reserves of knowledge; and it affords unlimited opportunities for that expertness in elementary techniques"\cite{Hardy1929}. Among the various results reported in his \emph{prolegomena}, one concerns the arithmetic and geometric means of a positive, bounded and Riemann integrable function $f(x)$ over an arbitrary interval $(a,b)$,
\begin{align}
\textgoth{U}(f)=\frac{1}{b-a}\int_{a}^{b} f(x) dx \;\;\;\textrm{and}\;\;\;
\textgoth{G}(f)=\frac{1}{b-a}\exp\left(\int_{a}^{b} \log{f} dx\right)\;\;\;\textrm{respectively.}
\end{align}
The theorem stating that $\textgoth{G}(f)\leq\textgoth{U}(f)$, however, is said to be ``not too easy to prove'', and a related result that $\textgoth{G}(f)=\lim_{r\rightarrow0}M_r(f)$ for the \emph{generalized arithmetic mean}
\begin{equation}
M_r(f)=\left(\frac{1}{b-a}\int_{a}^{b} f^r dx\right)^{1/r},
\end{equation}
``also presents some points of genuine difficulty''. In fact, ``there is no general proof [of it] in any book'', he added. However, in a subsequent letter to Hardy, Hungarian mathematician Fr\'ed\'eric (Frigyes) Riesz (1880-1956) showed that a simple trick of his finding can surmount both hurdles. From the identity $\lim_{n\rightarrow0} (f^n-1)/n=\log f$, albeit for $n\in\mathbb{R}$, it indeed follows that
\begin{equation}
\lim_{n\rightarrow0} [\textgoth{U}(f^n)-1]/n=\textgoth{U}(\log f)=\log \textgoth{G}(f),
\end{equation}
and the two theorems can then be straightforwardly demonstrated. 

Over the five years that followed Hardy's presidential address, the \emph{chapter}\footnote{Hardy initially considered submitting the work to \emph{Cambridge Tracts in Mathematics and Mathematical Physics}~\cite[p.~v]{hardy1934}, a series to which he had previously contributed a few fascicles~\cite{hardy1905,hardy1910,hardy1915}.} grew into a full-scale monograph, co-authored with his long-time collaborator, J.~E. Littlewood (1885-1977), and a regular Cambridge visitor, George P\'olya (1887-1985)~\cite{hardy1934,rice2003}. That book, which laid the basis for the modern study of inequalities~\cite{fink2000}, is still in print. Its chapter on integrals includes Riesz' demonstration of the original results---now part of Theorem 187---generalized to means weighed by probability distribution~\cite[Eq.~(6.8.3)]{hardy1934}.

\section{Early Replica Trick in Physics: Kac, Edwards and Ma}
Despite the mathematical elegance and usefulness of Reisz's identity, it does not appear to have been known by the statistical physicists who later used it as the basis of their \emph{replica trick}. (None of the mathematicians who have since used Theorem 187 seem aware of its physical relevance either; see, e.g.,~\cite{love1986,luor2010}.) The physics tradition can instead be traced back to three seemingly independent reformulations of the result around 1970.

It is not averaging over the free energy, but over the density of state of a chain of harmonic oscillators with random masses that brought mathematical physicist Mark Kac (1914-1984) to the replica trick. (Fittingly, Freeman Dyson (1923-2020) first proposed that model, motivated by ``a question of [Charles] Kittel [1916-2019], who was concerned with the thermal properties of glass''~\cite{dyson1953}, a system for which the RSB technique has since been particularly successful~\cite{parisi2020}.) Although Kac had once done formal work on the spectrum of analogous matrices~\cite{kac1953}, by the late 1960s he had long since moved on. His interest was nevertheless rekindled by the publication of a series of reprints---including Dyson's---about one-dimensional systems~\cite{mattis1966}. In a 1968 preprint based on a lecture given in Trondheim~\cite{kac1968}, where Kac was on sabbatical leave from the Rockefeller University~\cite[p.~146]{kac1985}, he proposed a replica trick to compute the logarithm of the determinantal equation, from which it is then possible to compute the average of the density of states. Although the proposed computation can only be done for integer powers of the quantity of interest, Kac made the ``extremely reasonable'' conjecture to analytically continue the expression to positive real numbers. Despite Kac's confidence in the scheme, he found ``considerable interest to investigate the relationship between [his] approach and the approach of Dyson''~\cite{kac1968}. One of postdocs, Ta-Feng Lin, soon took care of validating the equivalence~\cite{lin1970}. Kac is not a co-author of the subsequent paper, but given that he is acknowledged as having discussed the problem and helped writing the ensuing manuscript, the keenness of his interest cannot be denied. Once that particular check was complete, however, neither Lin nor Kac ever again paid attention to this mathematically unorthodox approach.

Given the difference in presentation style and the absence of any references to prior works, Manchester physicist Sam Edwards (1928-2015) had likely not heard Kac's talk nor seen his preprint when he started working on rubber\footnote{Even when Edwards worked out the average spectrum of symmetric random matrices using the replica trick a few years later, he did not cite Kac's use of that same identity~\cite{edwards1976}. Recent presentations of this topic also neglect Kac's prior work. See, e.g.,~\cite[Chap.~16]{livan2018} and \cite[Chap.~13]{bouchaud2020}.}. The stylistic contrast is nevertheless interesting. Edwards, following his advances on self-avoiding walks and polymer solutions, was encouraged by his Manchester colleagues Geoffrey Allen and Geoffrey Gee (1910-1996)~\cite{allen1999}, to obtain a ``semi-microscopic theory which, in modern parlance, described the universal properties of rubber that emerged from any polymer system''~\cite{goldbart2004}. Rubber, having crosslinks that are randomly distributed yet fixed for a given sample, precisely follows Brout's rationale for averaging over different realizations of disorder. For Edwards, considering the free energy $F(n)$ of a model with $n$ additional copies, and expanding for small $n$ then provided the average of interest. Although the approach implicitly requires analytically continuing integers to the reals, Edwards---unlike Kac---showed no qualms about it. He simply states: (emphasis added) ``suppose (\emph{as does indeed occur}) that $F(n)$ can be expanded''~\cite{edwards1971}, and that the mathematical assumption is ``\emph{a posteriori} justifiable''~\cite{edwards1972}. Edwards, understanding that the problem ``formulation is \emph{rigorously} founded'', assigned the project to a new graduate student, Rowan Deam~\cite{Deam1975}. Only five years later---following Edwards' move to Cambridge and his chairing of the Science Research Council (SRC)~\cite{warner2017}---was the completed effort published~\cite{deam1976}. Despite all that maturation, however, the analytic continuation in $n$ was still then not much of a concern to him.

In yet another independent effort, Shang-keng Ma (1940-1983) used a version of the replica trick to average over disorder in his 1972 preprint about an electron moving in a random potential. At the time, Ma was on sabbatical leave at Cornell, where he learned about Wilson's recent work on the renormalization group\footnote{``Remembering Shang-keng Ma,'' Shang-keng Ma Papers, Special Collections \& Archives, UC San Diego, Box 1, Folder 1}. That exposure led him to consider a $1/n$ expansion in the number $n$ of components of a vectorial field theory to capture the critical behavior of a Bose gas~\cite{ma1972}. In this context, it is not surprising that Ma might have also considered the zero-component limit, $n=0$, as a means to study disordered systems. Because the original preprint was never published (and no copy of it has yet been unearthed), his precise motivations remain somewhat elusive, but some of the work's content can nevertheless be reconstructed through citations\footnote{We note, in particular, that Grinstein's PhD thesis mentions the treatment of an electron in a random potential~\cite[p.~121]{grinstein1974}, and Ref.~\cite{ferrell1972} refers to an expansion in $n^{-1}$. Reference~\cite{rivier1977} points instead to Ma's monograph on critical phenomena~\cite[p.~413]{ma1976}, which mentions that ``the self-attracting walk problem [\ldots] is also equivalent to the problem of electronic motion in a random potential''. The former was solved by de Gennes~\cite{degennes1972}, and one might surmise that Ma pulled his preprint after noticing that equivalence.}. 
For our purposes, the most relevant of these is Geoffrey Grinstein's Harvard PhD thesis~\cite{grinstein1974}, obtained under Alan Luther's supervision. Grinstein indeed used ``a formal mathematical trick, devised by Ma,'' to formulate an effective Hamiltonian for an Ising model with random nonmagnetic impurities~\cite[p.~18]{grinstein1974}, which he derived ``order by order in perturbation theory, and [\ldots] used this to obtain critical exponents for a number of random models by using an expansion in powers of ($4-d$)''~\cite{emery1975}\footnote{The analysis of Grinstein's result by Victor Emery (1934-2002) in Ref.~\cite{emery1975} is what the ensuing publication cites~\cite{grinstein1976}.}.

By 1974, the replica trick was therefore starting to circulate in various theoretical physics circles. Given its relatively insouciant use by Edwards and others, had it not been for the burgeoning interest in a novel type of magnetic alloys, spin glasses, the underlying lack of mathematical rigor might not have raised much physical concern for some time longer.

\section{Replica Trick and Spin Glasses}
Anderson paid attention to the experimental work on spin glasses from its early days~\cite{anderson1970,anderson1972,anderson1972anomalous,anderson2004}. This interest naturally followed from him having worked on the Kondo effect, which was studied with the same alloys albeit at a lower concentration of magnetic impurities. His curiosity got further piqued when experimental results by John Mydosh and his graduate student at Wayne State University, Vincent Cannella, suggested that a non-standard transition might be at play in these systems~\cite{cannella1972,anderson2004,mydosh2021}. Spin glasses therefore putatively exhibited a different type of physics, more complex than what was found in ordered systems, with potentially ``many other applications in disordered state physics''~\cite{edwards1975}\footnote{Anderson's biographer, however, found no particular association between complexity and his early interest in spin glasses.~\cite[p.~260]{zangwill2021}.}. In 1974, the time was ripe for Anderson to dive into the topic. He was  dissatisfied~\cite{sherrington2020} with a recent theoretical proposal that ascribed the physics of spin glasses to a distribution of internal magnetic fields~\cite{adkins1974thesis,adkins1974}, and Edwards, his new Cambridge colleague, had just asked him for a problem on which to work during the long commute back and forth from the SRC in London.

When Edwards heard the spin glass proposal from Anderson, he immediately saw the opportunity to adapt his replica approach to this other type of quenched disorder. Of this interaction, Anderson later recalled: 
\begin{quote}
Sam mentioned to me that he happened to have in his notebook a method he had run into in the problem of gels, for which it wasn't well suited--but he thought this was the perfect instance. In a couple of weeks, at most, he brought in the replica solution which constitutes the main body of that paper. For me at least, and I think also for Sam, the self-consistent mean field technique of the earlier part of the paper was a welcome check on the very unfamiliar mathematics of the replica method, mathematics in which at every stage convergence seemed problematical until it happened~\cite{anderson2004}. 
\end{quote}
A series of joint publications on spin glasses introduced the Edwards-Anderson (EA) model,
\begin{equation}
\mathcal{H}^{(\mathrm{EA})}=\sum_{\langle ij\rangle} J_{ij} \mathbf{s}_i\cdot \mathbf{s}_j
\end{equation}
for nearest-neighbor classical Heisenberg spins, $\mathbf{s}_i\in S^2$, on a cubic lattice, with coupling constants $J_{ij}$ taken at random from a Gaussian distribution. (In other words, nearest-neighbor sites randomly interact through ferromagnetic or antiferromagnetic couplings, which frustrate the formation of ordered states at low temperatures.) The first of these papers, in particular, presented the replica trick solution of the EA model~\cite{edwards1975}.

Despite Edwards' busy administrative schedule in London, he tried to remain active as a mentor as well. He notably asked one of his Cambridge students to relocate to Imperial College London, so as to be physically closer~\cite{sherrington2020,warner1977}. During Edwards' visits, he would also discuss with his former student, then on the faculty, David Sherrington. Sherrington recall that Edwards: ``wanted someone else to talk to about [spin glasses]. He was in London and I was in London, and we knew one another, we got on rather well and we thought in similar fashions, and so he'd talk to me''~\cite{sherrington2020}. By that time, Sherrington was already working on disordered magnetic materials~\cite{sherrington1974}, in part thanks to his experimental colleague Brian Coles (1926-1997), who was an early advocate for the study of spin glasses and is even credited with coining the term~\cite{coles1999}\footnote{Terminological primacy is debated. In an early report, Anderson credits Coles for the term ``magnetic glasses''~\cite{anderson1970}, which he himself shortened to ``spin glasses''~\cite{anderson1992}.}.

Upon hearing about Edwards' work, Sherrington quickly tried a couple of ideas~\cite{sherrington1975b,sherrington1975c,southern2021}, before honing in on wanting ``to find something for which I could apply a kind of test of other approximations Sam was making, by looking at a problem which should be exactly solvable. I had enough background to know what that might be''~\cite{sherrington2020}. That background was knowing that a mean-field theory treatment should be exact for fully-connected models, in which each spin interacts with all others equally. Another of Sherrington's choice was more fortuitous. While Edwards---motivated by experiments---considered Heisenberg spins, Sherrington opted for Ising spins, $s_i=\pm1$, which seemed algebraically simpler~\cite{sherrington2020}, thus giving
\begin{equation}
\mathcal{H}^{(\mathrm{SK})}=\sum_{ij} (J_0+J_{ij}) s_i s_j,
\end{equation}
for a $J_0$ ferromagnetic offset (Fig.~\ref{fig:SKPD}).

It took some time for the consequences of that choice to sink in. Although Sherrington's derivation using replicas was complete by the spring of 1975, it is only during a sabbatical leave that following fall at IBM's Watson Research Center, in Yorktown Heights, that he got to collaborate with Scott Kirkpatrick. Kirkpatrick, who had worked on magnetism and percolation, quickly got interested in this new problem. His computational versatility further led him to consider numerical solutions to Sherrington's equations~\cite{kirkpatrick2021}. The two were then ``able to recognize that even though many of the results looked right, there was a serious problem, namely that the entropy was negative at $T=0$, which cannot occur for discrete spins,'' such as Ising spins. That finding was sufficiently surprising that by mid-October already the SK manuscript was at \emph{Physical Review Letters}~\cite{sherrington1975}.

\section{Towards Breaking RS: Thouless et al.}
\begin{figure}
\includegraphics[width=0.45\textwidth]{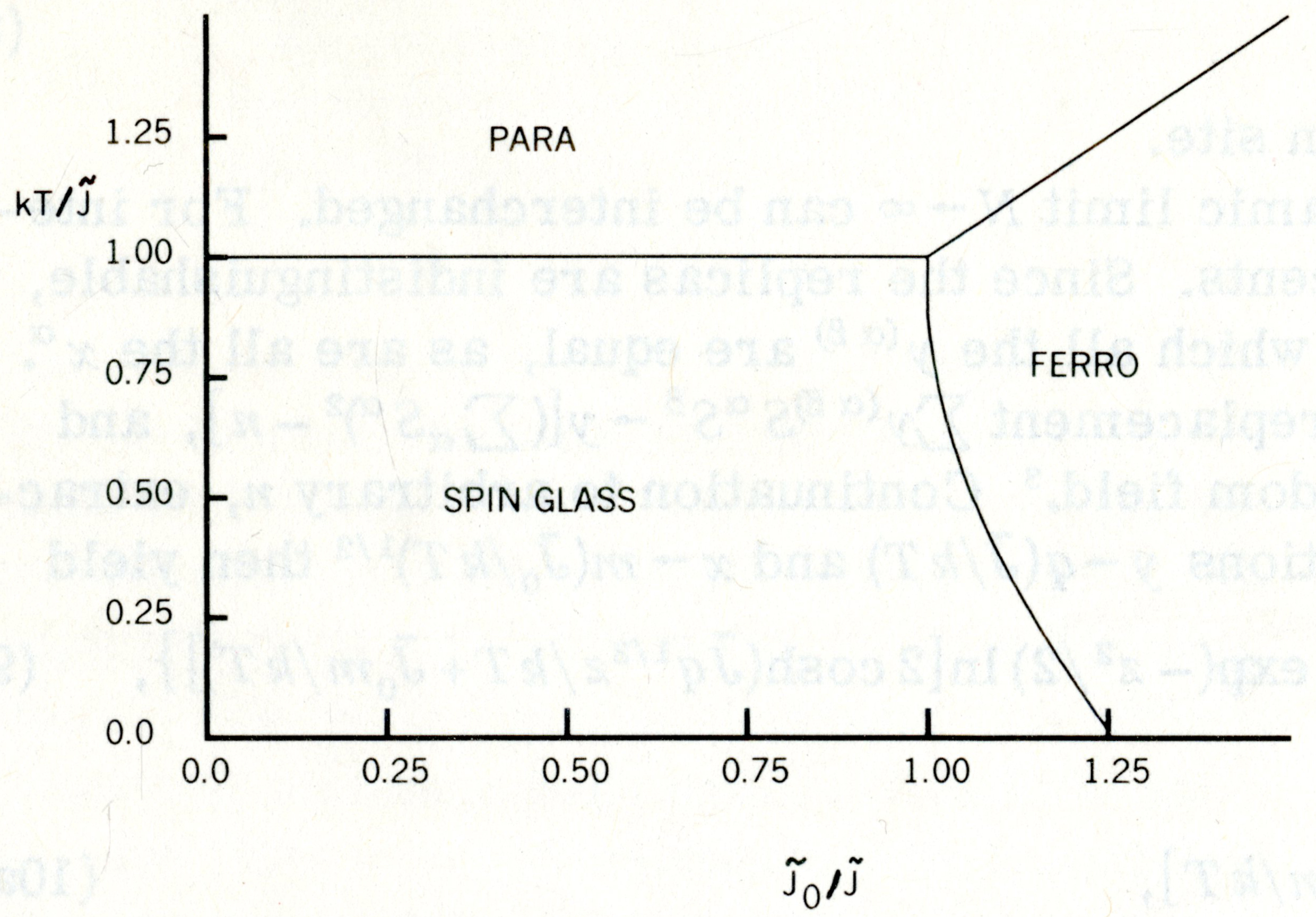}\includegraphics[width=0.45\textwidth]{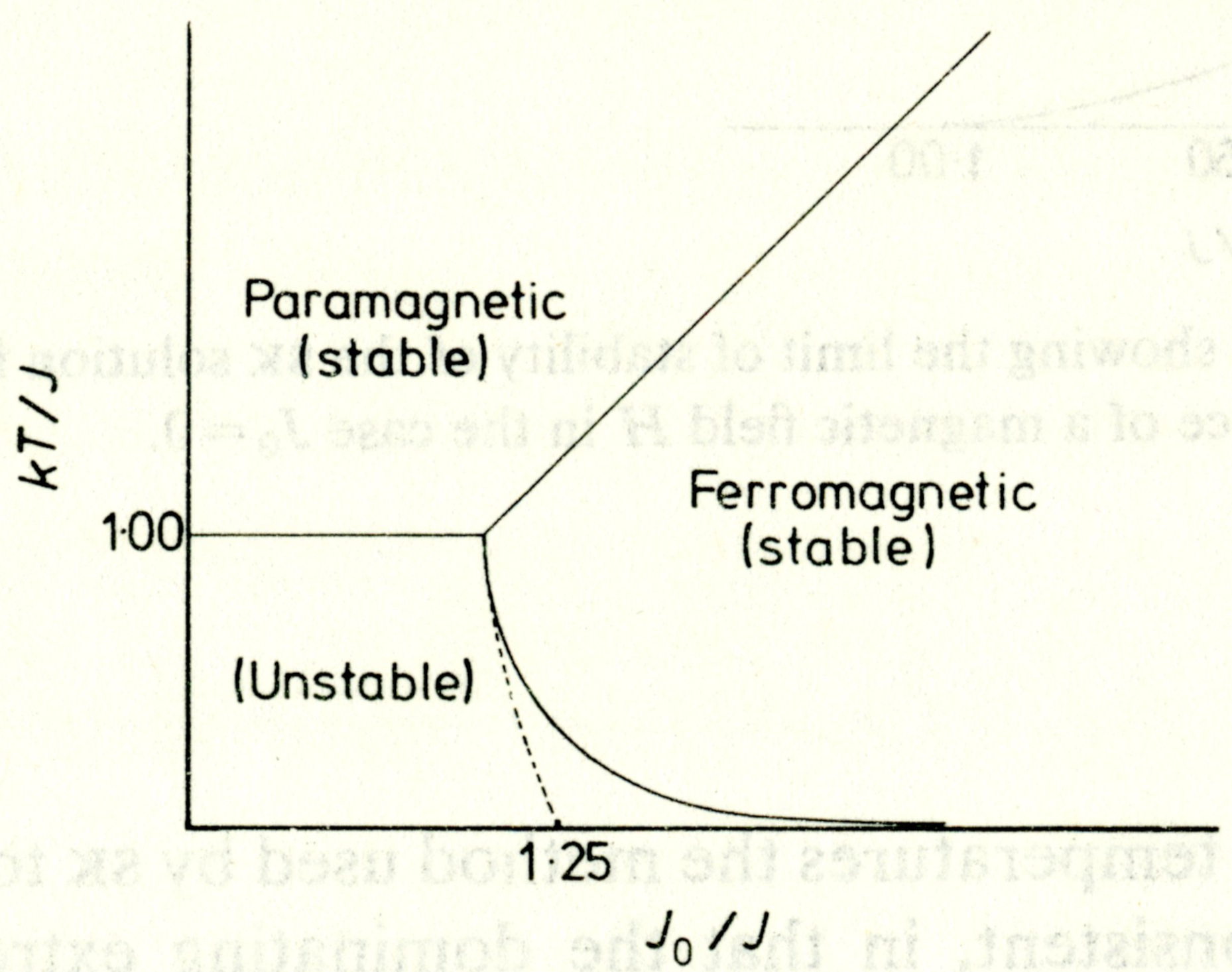}
\caption{Phase diagram of the SK model as a function of temperature $T$ and ferromagnetic bias $J_0$ both rescaled by the variance of $J_{ij}$: (left) The original result~\cite{sherrington1975}, and (right) identifying the replica instability region of de Almeida and Thouless~\cite{dealmeida1978}. (left) Reprinted figure with permission from D.~Sherrington and S.~Kirkpatrick. Phys.~Rev.~Lett. 35, 1792 (1975) \copyright (1975) by the American Physical Society. (right) Taken from Ref.~\cite{dealmeida1978} \copyright(1978) IOP Publishing.  Reproduced with permission.  All rights reserved.}
\label{fig:SKPD}
\end{figure}

As a referee of the SK manuscript, Birmingham physicist David Thouless (1934-2019) got an early peek at the physical inconsistency of its results~\cite{dealmeida2021,sherrington2020}. Sherrington recalls Thouless---as reviewer---insisting that: ``It can't be exactly solvable, because it's wrong!" He added: ``Of course, it's the solution that wasn't exact. The model itself was solvable. [Thouless] accepted `solvable', he wouldn't accept `exactly solvable',''\cite{sherrington2020}. Michael Moore, who joined Manchester after Edwards' departure and regularly ran into Thouless at that time, remembers that ``Thouless' original hypothesis was that Sherrington had simply cocked up the calculation completely, but then he did the calculation himself and he discovered it was perfectly ok'' \cite{moore2020}. In fact, Thouless had the calculation checked ``by more careful people than [him]''\footnote{Letter from Thouless to Edwards, December 15, 1975, Thouless Papers, Royal Society, Box 6.}, but Moore's point persists. 

Interestingly, it was not the replica approach that bothered Thouless. As he wrote to Edwards, ``[it] is not so much the $n=0$ business which worries me, as I am learning to accept that.'' In a separate letter to Anderson a few days earlier, he wrote more explicitly: ``I suspect the error in the SK treatment comes in the assumption that the main contribution to the steepest descents integral comes from a point at which all the $n$ values of $x^{\alpha}$ are equal and all the $n (n -1)$ values of $y^{\alpha\beta}$ are equal. Hard to argue about this when $n$ is zero, however''\footnote{Letter from Thouless to Anderson, 25 November 1975, J.~R.~L.~de Almeida personal collection.}. In this excerpt, Thouless explains that the problem is that all $n(n-1)/2$ pairs of replicas $\alpha\beta$ (with $\alpha\neq\beta$), which end up coupled as part of the calculation of $\overline{Z^n}$, are treated equivalently. A possibility might be that different groups of replicas behave differently, thus breaking this \emph{symmetry between replicas}. Pursuing this direction, however, was intricate. Moore also remembers Thouless explaining that ``[t]here must be an infinite number of ways to break replica symmetry. What are the chances that one hits on the right scheme? With an $n \times n$ matrix, you can parameterize it in endless ways, then you start fiddling with it. What criteria would you use to choose a solution?''

Given these early insights, it was natural for Thouless to study the stability of the symmetric assumption for replicas. By July 1977, he had already done part of the calculation, when he assigned the work to his recently-arrived Brazilian graduate student, Jairo de Almeida~\cite{dealmeida2021}\footnote{See also ``Research notebook $\sim1977$,'' Thouless Papers, Royal Society, Box 02.}. Thouless told him up front: ``The problem is with the replica symmetry thing that [SK] are doing. If you do the stability analysis, then it is going to be negative.'' The work, which was extended to systems with a ferromagnetic bias ($J_0>0$) by de Almeida, was submitted for publication later that fall (Fig.~\ref{fig:SKPD})~\cite{dealmeida1978}. It would open the door to studies of replica symmetry breaking, as other possible sources of discrepancy were gradually eliminated\footnote{At Duke, Richard Palmer and Leo Van Hemmen, then both young faculty members, considered the impact of inverting the $n\rightarrow0$ and the thermodynamic $N\rightarrow\infty$ limits in the saddle-point evaluation, but found no problem with it~\cite{vanhemmen1979,vanhemmen2021}.}.

Thouless, however, did not pursue this direction further. Before and after this result, he toiled instead on circumventing the difficulties of the replica scheme altogether. Already at the start of 1976, he was exchanging with Kirkpatrick, Anderson and his former student, Richard Palmer, as well as with Eliott Lieb about another approach\footnote{Thouless Papers, Royal Society, Box 12.}. A preprint about the ``Solution of 'Solvable model of spin glass' '' signed by the last four authors was then circulated. What later became known as the TAP paper further developed at the summer 1976 Aspen meeting on ``Current topics in the theory of condensed matter,'' and submitted that fall with a diminished author list~\cite{thouless1977}\footnote{The process that led to the final authorship selection of this work remains a bit murky; see, e.g., Ref.~\cite{kirkpatrick2021}.}. Thouless' other original works on spin glasses mostly sought out a mean-field solution for a finite-connectivity Cayley tree (Bethe lattice). This effort, which he had started in 1977 already\footnote{``Research notebook $\sim$ 1977,'' Thouless Papers, Royal Society, Box 02}, only reached publication a decade or so later~\cite{thouless1986,chayes1986,carlson1988}.

Others did not similarly equivocate. Moore, who first heard about the SK model from Michael Kosterlitz, Thouless' close collaborator (see, e.g., Ref.~\cite{kosterlitz1976}), also developed an interest in the topic. Despite largely buying into Thouless' skepticism about the chances of success of RSB, he and his Manchester colleague, Alan Bray, attacked the replica problem head on~\cite{bray1978}. ``We bashed away at this two-group method of breaking replica symmetry, which divided the replicas into two groups: $m$ in one category, $n - m$ in the other. It seemed to be the simplest scheme one could think of, but much to our astonishment we could get a stable solution.'' Their scheme, however, turned out to be fundamentally flawed in that $\lim_{n\rightarrow0} \overline{Z^n(m)}\neq 1$~\cite{moore2020}. 

\section{Actually Breaking RS: Blandin}
A completely separate attempt to break replica symmetry was undertaken by Andr\'e Blandin (1933-1983). Blandin had gotten interested in magnetic systems with random couplings during his thesis days, as one of Jacques Friedel's (1921-2014) first students at l'Universit\'e de Paris, in Orsay. Interestingly, he then entertained that randomly distributed manganese impurities in copper would give rise to an effective Hamiltonian
\begin{equation}
\mathcal{H}^{(\mathrm{B})}=-\sum_{\substack{i,j\\ \mathrm{Mn\;atoms}}} f(r_{ij}) \mathbf{s}_i\cdot \mathbf{s}_j
\end{equation}
with oscillatory couplings $f(r_{ij})$ of the Ruderman-Kittel-Kasuya-Yosida (RKKY) type, but ``à faible concentration, nous pouvons négliger la structure cristalline et considérer les atomes répartis au hasard comme dans un liquide''~\cite[p.~59-60]{blandin1961}. In other words, the couplings should be considered as randomly distributed, thus making the model akin to that later proposed by Edwards and Anderson. At low temperatures, Blandin expected this model to exhibit ``un antiferromagnétisme o\`u
chaque spin est fix\'e, mais \`a des positions au hasard dans l'espace''~\cite[p.~60]{blandin1961}. (For this antiferromagnetism with random-yet-fixed spin positions, Friedel would later prefer the term ``le désordre magnétique gelé'' over ``spin glass''~\cite[p.~203]{friedel1994}.) Blandin, however, added that further study of this model ``dépasse le cadre de ce travail''~\cite[p.~63]{blandin1961}. This (largely unpublished\footnote{According to Friedel~\cite[p.~203-204]{friedel1994}, Blandin discussed some of these ideas at a meeting in Oxford with Walter Marshall (1932-1996), who then published some of this material~\cite{marshall1960}, without properly crediting its origin. In that work, Marshall cites an earlier publication by Blandin and Friedel that only mentions some of these ideas~\cite{blandin1959}.}) analysis was indeed not pursued by Blandin for nearly two decades. During that time, he built a solid career as a theorist in Friedel's group, notably working on Anderson localization and the Kondo effect, thereby keeping a strong interest in the physics of systems with magnetic impurities~\cite{friedel1985}.

It is unclear whether Blandin followed the early experimental work on spin glasses, but by 1975 he was tracking closely that of Edwards and Anderson~\cite[p.~C6-1508]{blandin1978}. His first publication on the topic, which he presented at a low temperature physics conference in August 1978\footnote{XVth International Conference on Low Temperature Physics, Grenoble, France, August 23-29, 1978.}, carefully discusses prior theoretical efforts~\cite{blandin1978,toulouse1985}\footnote{During the summer of 1978, Blandin tried to get in touch with de Almeida to discuss his recent work with Thouless, but that meeting fell through~\cite{dealmeida2021}.}. More importantly, it proposes a clear scheme and rationale for breaking the symmetry between replicas. His subdividing replicas in $n/m$ groups with $m=2$ (i.e., in pairs) was motivated by the fact that the overlap $q$, which Edwards and Anderson understood as an order parameter for spin glasses, is a two-replica coupling field. Applying a (vanishing) coupling between these pairs should get them to share a same free energy minimum, while uncoupled replicas should end up distant in configuration space. 

Despite there being ``many ways of breaking the symmetry''~\cite[p. C6-1512]{blandin1978}, Blandin also had a vision for pursuing more generic coupling forms. A few months prior to that publication, he had recruited Thomas Garel and Marc Gabay as graduate students to work on this problem. Gabay recalls:
\begin{quote}
One day, it was a Sunday, the phone rang at my home. I picked up the phone and Blandin said: ``We've got to meet in this café in Montparnasse.'' [\ldots] Blandin started excitedly to jot down equations on a piece of paper and to elaborate on some ideas. [\ldots] In retrospect, of course, I realize that he had essentially the right idea, and now I fully understand what he was sort of thinking. [\ldots] Incrementally, indeed by working on more sophisticated schemes, we were able to find solutions that were getting better and better, or less bad and less bad.~\cite{gabay2021}
\end{quote}

Although this effort is cited as ``to be published'' in Blandin's article for the conference proceedings~\cite{blandin1978}, it was not rushed to a journal. The next manuscript, which teasingly mentions that ``one may question the validity of defining \emph{an} order parameter in the usual way, for this transition'' (emphasis in the original)~\cite{blandin1980}, was sent out only in late June of 1979. Blandin was on the right track, but his students doing the work were just getting started in this new research direction. Blandin had by then also been battling depression for over a decade~\cite{lubensky2021,kohn1990,gabay2021} and even had to be hospitalized at times. This plight, piled on his teaching responsibilities as professor, limited the extent of his personal engagement with the work. Moreover, he likely did not see anyone else toiling in this particular direction, and therefore might not have seen any particular urgency with advancing his RSB program.

\section{Full RSB: Parisi}
\begin{figure}
\includegraphics[width=0.45\textwidth]{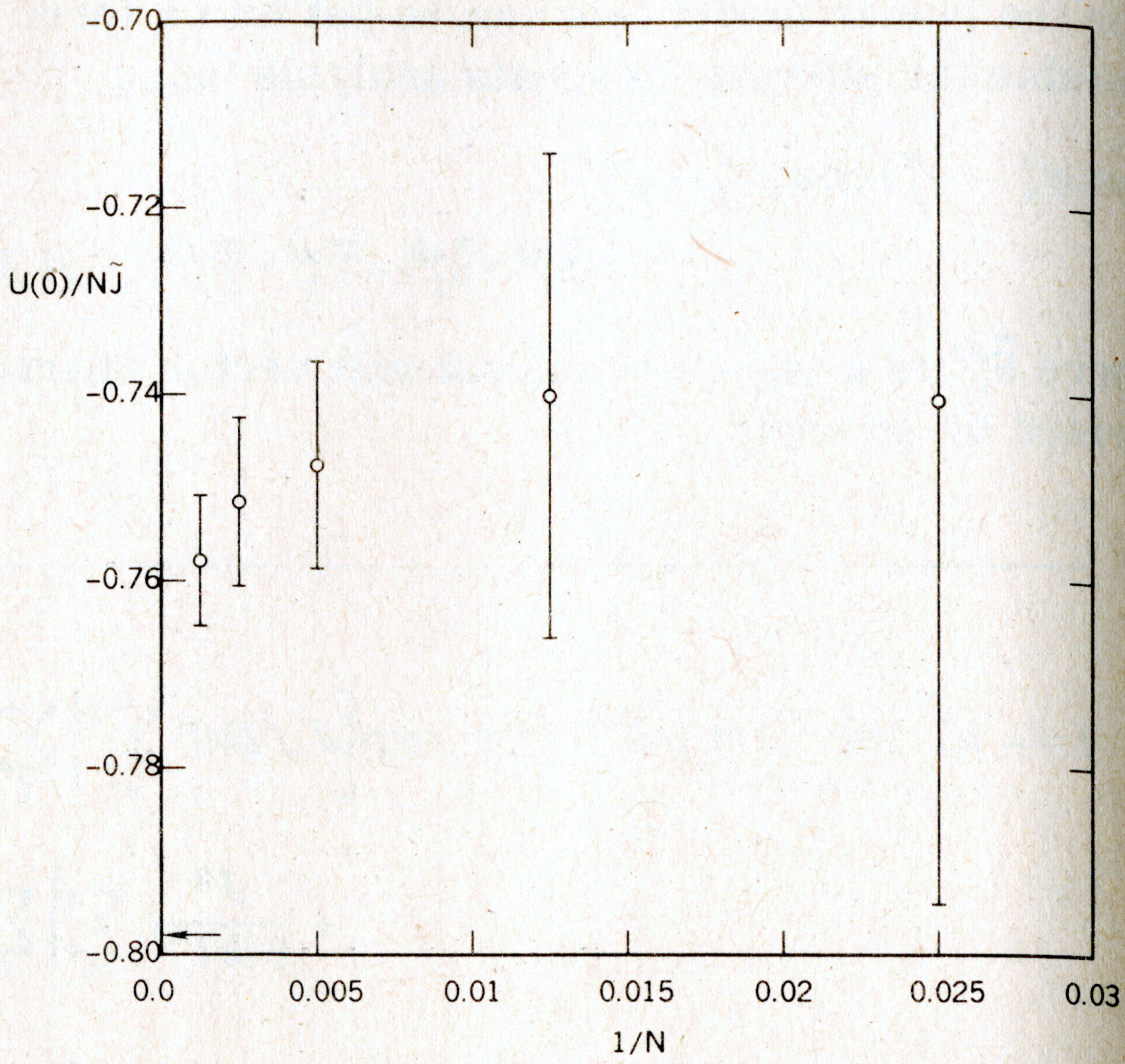}\includegraphics[width=0.45\textwidth]{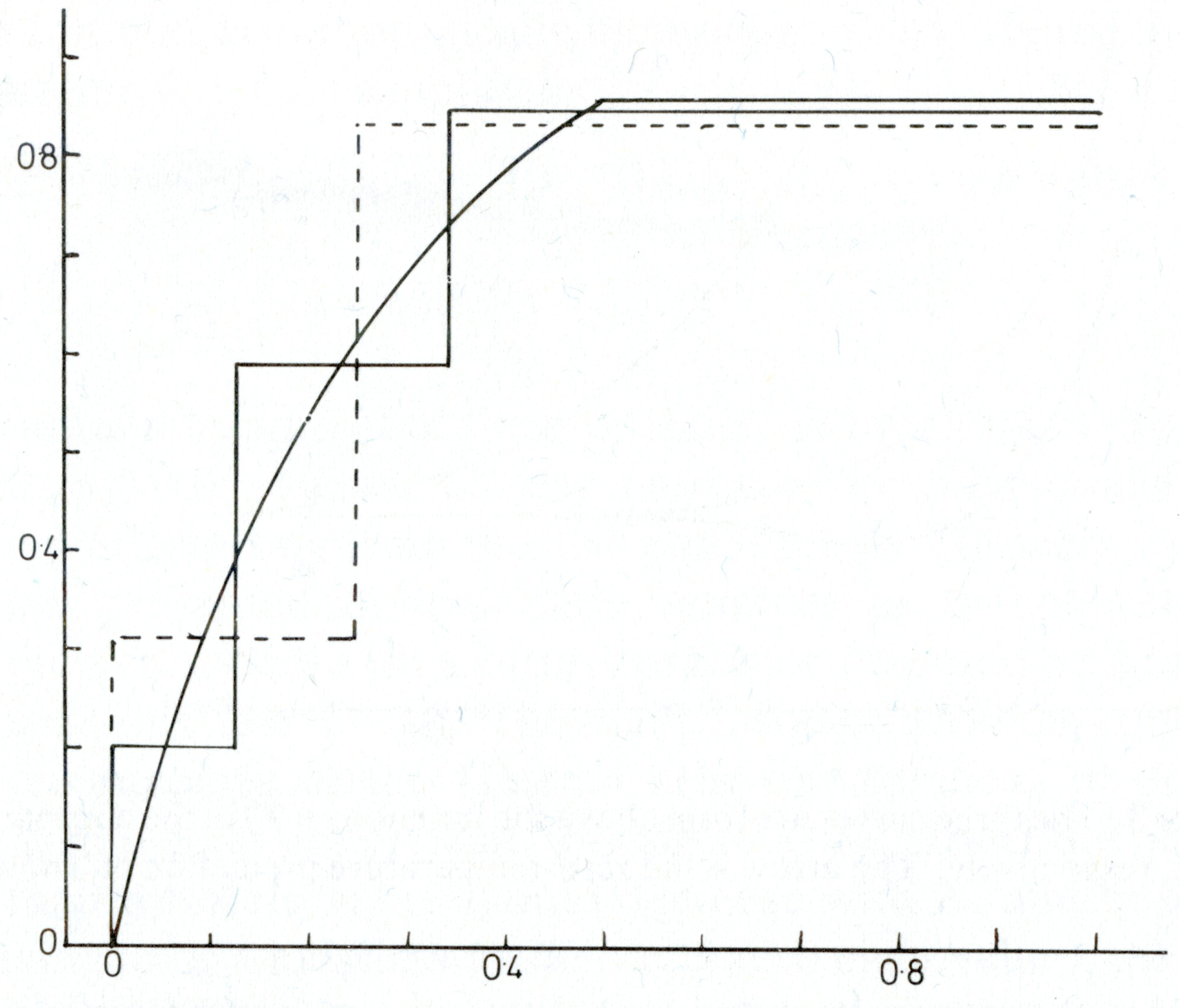}
\caption{(left) Ground state energy $U(0)$ per spin of the SK model with $J_0=0$ as a function of system size $N$ as computed by minimization. The simulation value tends toward -0.76(1) for $N\rightarrow\infty$. The arrow indicates the RS energy in that same limit~\cite{kirkpatrick1978}. (right) Overlap between replicas, $q(x)$ with $x=m$, for the SK model using the $k$RSB scheme with $k=1$, 2, and $\infty$, as obtained from the Parisi equation~\cite{parisi1980}. (left) Reprinted figure with permission from S.~Kirkpatrick and D.~Sherrington, Phys.~Rev.~B 17, 4384 (1978) \copyright(1978) by the American Physical Society. (right) Taken from Ref.~\cite{parisi1980} \copyright(1980) IOP Publishing.  Reproduced with permission.  All rights reserved.}
\label{fig:KSqx}
\end{figure}
During the summer of 1978, the Les Houches summer school in theoretical physics held a session on \emph{Ill-Condensed Matter}~\cite{leshouches1978}. This seminal get-together brought many of the key players in spin glass theory---Anderson, Kirkpatrick, Sherrington, and Thouless---along with a whole new generation of researchers to the French Alps. Given the growing theoretical and experimental enthusiams for spin glasses, the topic naturally appeared in various presentations. Anderson, in particular, gave a broad overview of the physics of amorphous systems that included a review of the problem with replicas, and specifically the ``very dubious and delicate mathematical extrapolation from finite integer $n$ to $n\rightarrow0$.'' He added that "[t]his extrapolation has already led us into deep, but not irrevocable, trouble even in exactly soluble cases''~\cite{anderson1978}. For those who attended the school, like Bernard Derrida, this presentation was a source of motivation to dive into the problem~\cite{derrida2020}; for those who were not there, a preprint of Anderson's notes quickly circulated. (See, e.g.,~\cite[Ref.~34]{blandin1978} and \cite[Ref.~6]{chaudhari1979}.) 

One of those who got a hold of that preprint is Giorgio Parisi, then a research scientist at the Istituto Nazionale di Fisica Nucleare in Frascati, outside of Rome. By a strange turn of events, Parisi was particularly receptive to this material. Up until that time, Parisi had not given much attention to condensed matter or statistical physics. Although he did work on critical exponents with some of his close university friends in the early 1970s~\cite{eramo1971,parisi2022}, he had quickly turned his attention to high energy physics. In 1978, in particular, he was trying to understand the importance of excitations in lattice gauge theory with Jean-Michel Drouffe and Nicolas Sourlas, whom he regularly visited in Paris. In that context, he surmised that ``you can have some excitation like a cube that is added on the surface. Other than this cube, [you] can add some other cube [\ldots] and there you can have some kind of polymer from the cube, which could have some bifurcations and so on.''~\cite{parisi2022} While surveying the literature on that theme, the team fell upon the recent work of Tom Lubensky and his UPenn graduate student, Joel Isaacson, on a field theory of branched polymers~\cite{lubensky1978}. Lubensky had learned about the replica trick from Luther~\cite{lubensky2021,lubensky1975}, and first used it to study the critical properties of spin glasses~\cite{lubensky1976}\footnote{Lubensky and his UPenn colleague, Brooks Harris, had also previously considered the impact of coupling disorder in a ferromagnet, although not in the spin glass limit, $J_0\rightarrow0$~\cite{harris1974,lubensky2021}.}. Hence, by the time of his branched polymer papers, taking $n\rightarrow0$ warranted but a brief technical mention.

Because Parisi was quite interested in this particular result, he nevertheless took note of the technique and cross-referenced it with Anderson's Les Houches notes. Despite having no interest in spin glasses, he then thought:
\begin{quote}
We cannot remain with something wrong that is written in the literature, that some method gives some wrong result and we don't know why. [\ldots] I think that I should read the literature and I think that it can be fixed easily. [\dots] I thought it should be fixed relatively easily.~\cite{parisi2022}
\end{quote}
This effort, which started around Christmas of 1978, took particular note of the numerical simulations that Kirkpatrick and Sherrington had published earlier that year. In addition to the original negative zero-temperature entropy result, there was ``good evidence in [that] second paper that the ground state energy was not $-0.798$ [as SK had computed], but was around $-0.76$'' (Fig.~\ref{fig:KSqx}). Parisi's early efforts led him to independently reproduce Blandin's  approach and to generalize it for any integer $m$, thus finding that ``the entropy at zero temperature is proportional to $1/m$, so the entropy problem is cured when $m$ goes to infinity, but the internal energy does not move.'' 

That particular result is what Parisi presented as a poster at a Trieste meeting, in March 1979\footnote{Middle European Cooperation in Statistical Physics (MECO) Sixth International Seminar on Phase Transitions and Critical Phenomena, March 26 – 28, 1979, International Center for Theoretical Physics, Trieste, Italy.}.  Imre Kondor, who crossed the Iron Curtain from Hungary to attend, recalls that ``that poster was a peculiar one. It was just two pages torn from a copy book with some scribbles on them.'' Even so, one of the attendees noted that this calculation ``does not make sense. You have a free energy and the free energy should be minimized. Now, you have [two] parameters [\ldots] and for one parameter you minimize and for the other parameter you maximize. [\ldots] You should minimize with respect to the two''~\cite{parisi2022}. Although the basis of that remark is now known to be erroneous in the limit $n\rightarrow\infty$, it got Parisi thinking that one should consider $m$ to be a real number and therefore the possibility that $m<1$, which would make both optimizations alike. In this case, the low-temperature limit is not algebraically solvable, but Parisi was undismayed. Given his prior coding experience and his access to a Cray-designed supercomputer, the CDC 7600, through a terminal at Frascati, he could surmount that hurdle. The result was that the entropy---albeit still negative---was much smaller and, crucially, that the energy was quite close the computational one. He recalls: ``I was extremely satisfied by this thing, because it was the first time that I could see something that more or less solved, or nearly solved, both problems.'' That April, he sent a manuscript out with this result~\cite{parisi1979}. The reviewer found that ``the construction is completely incomprehensible, but as long as the formula gives the correct result, the result goes in the right direction---the energy is correct and so on---the paper should be published''~\cite{parisi2022}. 

Like Blandin, Parisi speculated that ``it is quite likely that an infinite number of order parameters is needed in the correct treatment''\footnote{Following Moore's results, Peter Young had concluded the same~\cite{young1979,young2021}.}. Parisi, however, then promptly carried through with the idea of dividing replicas in a hierarchy of subgroups. He recalls:
\begin{quotation}
If you look from the point of view of symmetry, in essence what was broken was the $\mathrm{O}(n\rightarrow0)$ group. When you have this broken symmetry, you have still an $\mathrm{O}(0)$ group that remains unbroken, so you could break again that group [\ldots]. I was familiar---from high-energy physics, from my thesis work---with the idea that you have a group, a breaking of the symmetry group, Goldstone bosons and all this type of coset group that corresponds to the breaking. All the group theory was clear to me.~\cite{parisi2022}
\end{quotation}
However, even pursuing that idea was not mathematically trivial. Sourlas recalls that during the summer of '79\footnote{They were all attending the Carg\`ese Summer Institute: Recent Developments in Gauge Theories, 26 August-8 September 1979~\cite{cargese1980}.}: 
\begin{quotation}
Parisi himself was very much aware of the fact that his theory was mathematically unorthodox. I very vividly remember an after-dinner discussion in a Carg\`ese caf\'e between Parisi and the Harvard mathematician Raoul Bott [1923-2005]. Parisi asks Bott: ``Can you define a matrix with zero elements?'' Bott tried for a few minutes to find a rigorous definition. [\ldots] Bott was very puzzled and discouraged and this was the end of the discussion as I remember it.~\cite{sourlas2021}
\end{quotation}
Yet Parisi persisted.
\begin{quotation}
I started to do two-step, three-step near the critical temperature. That was a lot of complex computation because the algebra is painful when you're doing two steps, three steps, four steps. [\ldots] I had the intuition that when you go to $k=\infty$ the interval [between steps] becomes smaller and smaller and the function becomes a continuous function. What was extremely surprising is that, when you write the formula near the critical point for a continuous function, everything simplifies a lot. [\ldots] That was more or less the situation before the summer. [\ldots] At the end of '79, [as] I was finishing the study, [\ldots] I realized that one could have a compact expression for the free energy.~\cite{parisi2022}
\end{quotation}

Parisi's first series of works on RSB ended with this compact \emph{Parisi formula} solution the SK model (see Fig.~\ref{fig:KSqx}). Its ``zero-temperature entropy is consistent with zero, while the zero-temperature internal energy is estimated to be $-0.7633\pm 10^{-4}$''~\cite{parisi1980}, which agrees with Kirkpatrick's numerical results for the ground state. The reception of that series of results was overall quite good. After an after dinner talk at Les Houches in February of 1980, Parisi felt that: ``Most of the people present there were convinced by these things. There was a quite strong applause after the talk. I remember that Leo Kadanoff [1937-2015] was there, and I remember that he strongly congratulated me with the thing.'' Moore, who reviewed a few of these early manuscripts, also remembers that ``Giorgio’s scheme was developed quite quickly and was accepted very quickly as well''~\cite{moore2020}.

Of the subsequent steps, Parisi wrote: ``the computations of the fluctuations induced corrections, of the Goldstone modes and of the lower critical dimension are only technical problems which may be solved with a serious effort''~\cite{parisi1980reports}. Thouless, de Almeida and Kosterlitz were quick to check that his solution was at least stable around the transition temperature. 
\begin{quote}
This type of analysis [\ldots] was the final death knell of the original SK solution and any meaningful solution must survive such an analysis. [\ldots] Although this paper is further evidence that Parisi’s solution may be the correct one, it leaves the most important questions unanswered. For example, what is the physical meaning of the formal mathematical manipulations? [\ldots] How do we incorporate spatial fluctuations in $q$ and, of course, does a spin glass phase exist in realistic models?~\cite{thouless1980}
\end{quote}

\section{Conclusion}
Following Parisi's breakthrough with RSB, he initially left spin glasses to others. In his words: ``A problem that we could solve, well, it's a problem that we can solve. A problem that we cannot solve, that needs new ideas, that's certainly interesting''~\cite{parisi2022}. The program sketched by Parisi and by Thouless \emph{et al.} seemed to fall in that second category. It turned out, however, to be somewhat harder to carry out than expected. In some ways, it is still ongoing. Over the last four decades, Parisi has therefore regularly rejoined the many physicists who bring new ideas to the study of spin glasses.

In parallel, the replica trick has turned into a \emph{technique}, applied to a variety of problems far beyond spin glasses. Within only a few years, it had spread to the study of neural networks, optimization problems, prebiotic evolution, and more. This expansion led Parisi to co-author the book \emph{Spin Glass Theory and Beyond}\cite{mezard1987}, and Anderson to write a seven-part series on spin glasses for \emph{Physics Today} that included \emph{Spin glass as cornucopia}~\cite{anderson1989} and \emph{Spin Glass as Paradigm}~\cite{anderson1990}. A contemporary effort to summarize the advances since has swelled into a book manuscript with well over 30 chapters~\cite{farbeyond2023}. The success of the technique is undeniable.

While resolving the problem with the replica trick was one of these proverbial mathematical \emph{tours de force}, getting to it required more than mere algebraic versatility. A finely tuned mathematical risk taking was also needed. Both theoretical physics' cavalier attitude towards rigor and a close attention to numerical and physical constraints were key for reaching a sound result. A formal approach to RSB would not have gotten there. In fact, it still hasn't. 
From a mathematical physics viewpoint, the technique remain safely distant from rigor. Although Parisi's RSB results for the SK model are now formally known to hold~\cite{talagrand2010,talagrand2011}, the tension with the physical efficacy of the approach persists. Not every scientist is a peace with that, which likely made the Physics Nobel committee equivocate for some time.

\begin{acknowledgments}
I thank my collaborator on this project, Francesco Zamponi, for his help and support. I thank him as well as Martin Niss for their critical reading of early drafts of this work. I especially thank all the oral history interviewees, without whom this article would not have been possible. I acknowledge support from the Simons Foundation Grant No.~454937. \textbf{Community appeal:} If anyone has photographs, documents, or recollections relevant to the \emph{History of RSB Project}, please do not hesitate get in touch with me.
\end{acknowledgments}

\bibliography{ReplicaTrick}

\begin{thebibliography}{104}%
\makeatletter
\providecommand \@ifxundefined [1]{%
 \@ifx{#1\undefined}
}%
\providecommand \@ifnum [1]{%
 \ifnum #1\expandafter \@firstoftwo
 \else \expandafter \@secondoftwo
 \fi
}%
\providecommand \@ifx [1]{%
 \ifx #1\expandafter \@firstoftwo
 \else \expandafter \@secondoftwo
 \fi
}%
\providecommand \natexlab [1]{#1}%
\providecommand \enquote  [1]{``#1''}%
\providecommand \bibnamefont  [1]{#1}%
\providecommand \bibfnamefont [1]{#1}%
\providecommand \citenamefont [1]{#1}%
\providecommand \href@noop [0]{\@secondoftwo}%
\providecommand \href [0]{\begingroup \@sanitize@url \@href}%
\providecommand \@href[1]{\@@startlink{#1}\@@href}%
\providecommand \@@href[1]{\endgroup#1\@@endlink}%
\providecommand \@sanitize@url [0]{\catcode `\\12\catcode `\$12\catcode
  `\&12\catcode `\#12\catcode `\^12\catcode `\_12\catcode `\%12\relax}%
\providecommand \@@startlink[1]{}%
\providecommand \@@endlink[0]{}%
\providecommand \url  [0]{\begingroup\@sanitize@url \@url }%
\providecommand \@url [1]{\endgroup\@href {#1}{\urlprefix }}%
\providecommand \urlprefix  [0]{URL }%
\providecommand \Eprint [0]{\href }%
\providecommand \doibase [0]{http://dx.doi.org/}%
\providecommand \selectlanguage [0]{\@gobble}%
\providecommand \bibinfo  [0]{\@secondoftwo}%
\providecommand \bibfield  [0]{\@secondoftwo}%
\providecommand \translation [1]{[#1]}%
\providecommand \BibitemOpen [0]{}%
\providecommand \bibitemStop [0]{}%
\providecommand \bibitemNoStop [0]{.\EOS\space}%
\providecommand \EOS [0]{\spacefactor3000\relax}%
\providecommand \BibitemShut  [1]{\csname bibitem#1\endcsname}%
\let\auto@bib@innerbib\@empty
\bibitem [{\citenamefont {{The {N}obel Committee for
  Physics}}(2021)}]{nobel2021}%
  \BibitemOpen
  \bibfield  {author} {\bibinfo {author} {\bibnamefont {{The {N}obel Committee
  for Physics}}},\ }\href@noop {} {\emph {\bibinfo {title} {Scientific
  Background on the Nobel Prize in Physics 2021 “For Groundbreaking
  Contributions to our Understanding of Complex Physical Systems”}}}\
  (\bibinfo  {publisher} {The Royal Swedish Academy of Sciences},\ \bibinfo
  {year} {2021})\BibitemShut {NoStop}%
\bibitem [{\citenamefont {Brout}(1959)}]{brout1959}%
  \BibitemOpen
  \bibfield  {author} {\bibinfo {author} {\bibfnamefont {R.}~\bibnamefont
  {Brout}},\ }\bibfield  {title} {\enquote {\bibinfo {title} {Statistical
  mechanical theory of a random ferromagnetic system},}\ }\href {\doibase
  10.1103/PhysRev.115.824} {\bibfield  {journal} {\bibinfo  {journal} {Phys.
  Rev.}\ }\textbf {\bibinfo {volume} {115}},\ \bibinfo {pages} {824--835}
  (\bibinfo {year} {1959})}\BibitemShut {NoStop}%
\bibitem [{\citenamefont {Talagrand}(2010)}]{talagrand2010}%
  \BibitemOpen
  \bibfield  {author} {\bibinfo {author} {\bibfnamefont {M.}~\bibnamefont
  {Talagrand}},\ }\href@noop {} {\emph {\bibinfo {title} {Mean Field Models for
  Spin Glasses: Volume I: Basic Examples}}}\ (\bibinfo  {publisher}
  {Springer},\ \bibinfo {year} {2010})\BibitemShut {NoStop}%
\bibitem [{\citenamefont {Talagrand}(2011)}]{talagrand2011}%
  \BibitemOpen
  \bibfield  {author} {\bibinfo {author} {\bibfnamefont {M.}~\bibnamefont
  {Talagrand}},\ }\href@noop {} {\emph {\bibinfo {title} {Mean Field Models for
  Spin Glasses: Volume II: Advanced Replica-Symmetry and Low Temperature}}}\
  (\bibinfo  {publisher} {Springer},\ \bibinfo {year} {2011})\BibitemShut
  {NoStop}%
\bibitem [{\citenamefont {Charbonneau}\ \emph {et~al.}(2023)\citenamefont
  {Charbonneau}, \citenamefont {Marinari}, \citenamefont {M\'ezard},
  \citenamefont {Ricci-Tersenghi}, \citenamefont {Sicuro},\ and\ \citenamefont
  {Zamponi}}]{farbeyond2023}%
  \BibitemOpen
  \bibinfo {editor} {\bibfnamefont {P.}~\bibnamefont {Charbonneau}}, \bibinfo
  {editor} {\bibfnamefont {E.}~\bibnamefont {Marinari}}, \bibinfo {editor}
  {\bibfnamefont {M.}~\bibnamefont {M\'ezard}}, \bibinfo {editor}
  {\bibfnamefont {F.}~\bibnamefont {Ricci-Tersenghi}}, \bibinfo {editor}
  {\bibfnamefont {G.}~\bibnamefont {Sicuro}}, \ and\ \bibinfo {editor}
  {\bibfnamefont {F.}~\bibnamefont {Zamponi}},\ eds.,\ \href@noop {} {\emph
  {\bibinfo {title} {Spin Glass Theory and Far Beyond: Replica Symmetry
  Breaking after 40 Years}}}\ (\bibinfo  {publisher} {World Scientific
  Publishing},\ \bibinfo {year} {2023})\BibitemShut {NoStop}%
\bibitem [{\citenamefont {Charbonneau}\ and\ \citenamefont
  {Zamponi}(2022{\natexlab{a}})}]{caphes2022}%
  \BibitemOpen
  \bibfield  {author} {\bibinfo {author} {\bibfnamefont {P.}~\bibnamefont
  {Charbonneau}}\ and\ \bibinfo {author} {\bibfnamefont {F.}~\bibnamefont
  {Zamponi}},\ }\href@noop {} {\enquote {\bibinfo {title} {L’{H}istoire de la
  brisure des r\'epliques en physique/{T}he {H}istory of {R}eplica {S}ymmetry
  {B}reaking in {P}hysics},}\ }\bibinfo {howpublished}
  {\url{https://caphes.ens.fr/history-of-replica-symmetry-breaking-in-physics/}}
  (\bibinfo {year} {2022}{\natexlab{a}}),\ \bibinfo {note} {accessed:
  2022-08-23}\BibitemShut {NoStop}%
\bibitem [{\citenamefont {Hardy}(1929)}]{Hardy1929}%
  \BibitemOpen
  \bibfield  {author} {\bibinfo {author} {\bibfnamefont {G.~H.}\ \bibnamefont
  {Hardy}},\ }\bibfield  {title} {\enquote {\bibinfo {title} {{Prolegomena to a
  Chapter on Inequalities}},}\ }\href {\doibase 10.1112/jlms/s1-4.1.61}
  {\bibfield  {journal} {\bibinfo  {journal} {J. London Math. Soc.}\ }\textbf
  {\bibinfo {volume} {s1-4}},\ \bibinfo {pages} {61--78} (\bibinfo {year}
  {1929})}\BibitemShut {NoStop}%
\bibitem [{\citenamefont {Hardy}\ \emph {et~al.}(1934)\citenamefont {Hardy},
  \citenamefont {Littlewood},\ and\ \citenamefont {P{\'o}lya}}]{hardy1934}%
  \BibitemOpen
  \bibfield  {author} {\bibinfo {author} {\bibfnamefont {G.~H.}\ \bibnamefont
  {Hardy}}, \bibinfo {author} {\bibfnamefont {J.~E.}\ \bibnamefont
  {Littlewood}}, \ and\ \bibinfo {author} {\bibfnamefont {G.}~\bibnamefont
  {P{\'o}lya}},\ }\href@noop {} {\emph {\bibinfo {title} {Inequalities}}}\
  (\bibinfo  {publisher} {Cambridge University Press},\ \bibinfo {year}
  {1934})\BibitemShut {NoStop}%
\bibitem [{\citenamefont {Hardy}(1905)}]{hardy1905}%
  \BibitemOpen
  \bibfield  {author} {\bibinfo {author} {\bibfnamefont {G.~H.}\ \bibnamefont
  {Hardy}},\ }\href@noop {} {\emph {\bibinfo {title} {The Integration of
  Functions of a Single Variable}}},\ \bibinfo {series} {Cambridge tracts in
  mathematics and mathematical physics}, Vol.~\bibinfo {volume} {2}\ (\bibinfo
  {publisher} {Cambridge University Press},\ \bibinfo {year}
  {1905})\BibitemShut {NoStop}%
\bibitem [{\citenamefont {Hardy}(1910)}]{hardy1910}%
  \BibitemOpen
  \bibfield  {author} {\bibinfo {author} {\bibfnamefont {G.~H.}\ \bibnamefont
  {Hardy}},\ }\href@noop {} {\emph {\bibinfo {title} {Orders of Infinity: The
  'infinit{\"a}rcalc{\"u}l' of Paul Du Bois-Reymond}}},\ \bibinfo {series}
  {Cambridge tracts in mathematics and mathematical physics}, Vol.~\bibinfo
  {volume} {12}\ (\bibinfo  {publisher} {Cambridge University Press},\ \bibinfo
  {year} {1910})\BibitemShut {NoStop}%
\bibitem [{\citenamefont {Hardy}\ and\ \citenamefont
  {Riesz}(1915)}]{hardy1915}%
  \BibitemOpen
  \bibfield  {author} {\bibinfo {author} {\bibfnamefont {G.~H.}\ \bibnamefont
  {Hardy}}\ and\ \bibinfo {author} {\bibfnamefont {M.}~\bibnamefont {Riesz}},\
  }\href@noop {} {\emph {\bibinfo {title} {The General Theory of Dirichlet's
  Series}}},\ \bibinfo {series} {Cambridge tracts in mathematics and
  mathematical physics}, Vol.~\bibinfo {volume} {18}\ (\bibinfo  {publisher}
  {Cambridge University Press},\ \bibinfo {year} {1915})\BibitemShut {NoStop}%
\bibitem [{\citenamefont {Rice}\ and\ \citenamefont {Wilson}(2003)}]{rice2003}%
  \BibitemOpen
  \bibfield  {author} {\bibinfo {author} {\bibfnamefont {A.~C.}\ \bibnamefont
  {Rice}}\ and\ \bibinfo {author} {\bibfnamefont {R.~J.}\ \bibnamefont
  {Wilson}},\ }\bibfield  {title} {\enquote {\bibinfo {title} {The rise of
  {B}ritish analysis in the early 20th century: the role of {G. H.} {H}ardy and
  the {L}ondon {M}athematical {S}ociety},}\ }\href {\doibase
  10.1016/S0315-0860(03)00002-8} {\bibfield  {journal} {\bibinfo  {journal}
  {Hist. Math.}\ }\textbf {\bibinfo {volume} {30}},\ \bibinfo {pages}
  {173--194} (\bibinfo {year} {2003})}\BibitemShut {NoStop}%
\bibitem [{\citenamefont {Fink}(2000)}]{fink2000}%
  \BibitemOpen
  \bibfield  {author} {\bibinfo {author} {\bibfnamefont {A.~M.}\ \bibnamefont
  {Fink}},\ }\bibfield  {title} {\enquote {\bibinfo {title} {An essay on the
  history of inequalities},}\ }\href {\doibase 10.1006/jmaa.2000.6934}
  {\bibfield  {journal} {\bibinfo  {journal} {J. Math. Anal. Appl.}\ }\textbf
  {\bibinfo {volume} {249}},\ \bibinfo {pages} {118--134} (\bibinfo {year}
  {2000})}\BibitemShut {NoStop}%
\bibitem [{\citenamefont {Love}(1986)}]{love1986}%
  \BibitemOpen
  \bibfield  {author} {\bibinfo {author} {\bibfnamefont {E.~R.}\ \bibnamefont
  {Love}},\ }\bibfield  {title} {\enquote {\bibinfo {title} {Inequalities
  related to those of hardy and of cochran and lee},}\ }\href {\doibase
  10.1017/S0305004100064343} {\bibfield  {journal} {\bibinfo  {journal} {Math.
  Proc. Camb. Philos. Soc.}\ }\textbf {\bibinfo {volume} {99}},\ \bibinfo
  {pages} {395–408} (\bibinfo {year} {1986})}\BibitemShut {NoStop}%
\bibitem [{\citenamefont {Luor}(2010)}]{luor2010}%
  \BibitemOpen
  \bibfield  {author} {\bibinfo {author} {\bibfnamefont {D.-C.}\ \bibnamefont
  {Luor}},\ }\bibfield  {title} {\enquote {\bibinfo {title} {Multidimensional
  exponential inequalities with weights},}\ }\href {\doibase
  10.4153/CMB-2010-038-1} {\bibfield  {journal} {\bibinfo  {journal} {Can.
  Math. Bull.}\ }\textbf {\bibinfo {volume} {53}},\ \bibinfo {pages}
  {327–339} (\bibinfo {year} {2010})}\BibitemShut {NoStop}%
\bibitem [{\citenamefont {Dyson}(1953)}]{dyson1953}%
  \BibitemOpen
  \bibfield  {author} {\bibinfo {author} {\bibfnamefont {F.~J.}\ \bibnamefont
  {Dyson}},\ }\bibfield  {title} {\enquote {\bibinfo {title} {The dynamics of a
  disordered linear chain},}\ }\href {\doibase 10.1103/PhysRev.92.1331}
  {\bibfield  {journal} {\bibinfo  {journal} {Phys. Rev.}\ }\textbf {\bibinfo
  {volume} {92}},\ \bibinfo {pages} {1331--1338} (\bibinfo {year}
  {1953})}\BibitemShut {NoStop}%
\bibitem [{\citenamefont {Parisi}\ \emph {et~al.}(2020)\citenamefont {Parisi},
  \citenamefont {Urbani},\ and\ \citenamefont {Zamponi}}]{parisi2020}%
  \BibitemOpen
  \bibfield  {author} {\bibinfo {author} {\bibfnamefont {G.}~\bibnamefont
  {Parisi}}, \bibinfo {author} {\bibfnamefont {P.}~\bibnamefont {Urbani}}, \
  and\ \bibinfo {author} {\bibfnamefont {F.}~\bibnamefont {Zamponi}},\
  }\href@noop {} {\emph {\bibinfo {title} {Theory of {S}imple {G}lasses:
  {E}xact {S}olutions in {I}nfinite {D}imensions}}}\ (\bibinfo  {publisher}
  {Cambridge University Press},\ \bibinfo {year} {2020})\BibitemShut {NoStop}%
\bibitem [{\citenamefont {Kac}\ \emph {et~al.}(1953)\citenamefont {Kac},
  \citenamefont {Murdock},\ and\ \citenamefont {Szeg{\"o}}}]{kac1953}%
  \BibitemOpen
  \bibfield  {author} {\bibinfo {author} {\bibfnamefont {M.}~\bibnamefont
  {Kac}}, \bibinfo {author} {\bibfnamefont {W.~L.}\ \bibnamefont {Murdock}}, \
  and\ \bibinfo {author} {\bibfnamefont {G.}~\bibnamefont {Szeg{\"o}}},\
  }\bibfield  {title} {\enquote {\bibinfo {title} {On the eigen-values of
  certain {H}ermitian forms},}\ }\href@noop {} {\bibfield  {journal} {\bibinfo
  {journal} {J. Ration. Mech. Anal.}\ }\textbf {\bibinfo {volume} {2}},\
  \bibinfo {pages} {767--800} (\bibinfo {year} {1953})}\BibitemShut {NoStop}%
\bibitem [{\citenamefont {Lieb}\ and\ \citenamefont
  {Mattis}(1966)}]{mattis1966}%
  \BibitemOpen
  \bibfield  {author} {\bibinfo {author} {\bibfnamefont {E.~H.}\ \bibnamefont
  {Lieb}}\ and\ \bibinfo {author} {\bibfnamefont {D.~C.}\ \bibnamefont
  {Mattis}},\ }\href@noop {} {\emph {\bibinfo {title} {Mathematical Physics in
  One Dimension: Exactly Soluble Models of Interacting Particles}}}\ (\bibinfo
  {publisher} {Academic Press},\ \bibinfo {year} {1966})\BibitemShut {NoStop}%
\bibitem [{\citenamefont {Kac}(2013)}]{kac1968}%
  \BibitemOpen
  \bibfield  {author} {\bibinfo {author} {\bibfnamefont {M.}~\bibnamefont
  {Kac}},\ }\bibfield  {title} {\enquote {\bibinfo {title} {On certain
  toeplitz-like matrices and their relation to the problem of lattice
  vibrations},}\ }\href {\doibase 10.1007/s10955-012-0675-7} {\bibfield
  {journal} {\bibinfo  {journal} {J. Stat. Phys.}\ }\textbf {\bibinfo {volume}
  {151}},\ \bibinfo {pages} {785--795} (\bibinfo {year} {2013})}\BibitemShut
  {NoStop}%
\bibitem [{\citenamefont {Kac}(1985)}]{kac1985}%
  \BibitemOpen
  \bibfield  {author} {\bibinfo {author} {\bibfnamefont {M.}~\bibnamefont
  {Kac}},\ }\href@noop {} {\emph {\bibinfo {title} {Enigmas Of Chance}}}\
  (\bibinfo  {publisher} {Basic Books},\ \bibinfo {year} {1985})\BibitemShut
  {NoStop}%
\bibitem [{\citenamefont {Lin}(1970)}]{lin1970}%
  \BibitemOpen
  \bibfield  {author} {\bibinfo {author} {\bibfnamefont {T.‐F.}\ \bibnamefont
  {Lin}},\ }\bibfield  {title} {\enquote {\bibinfo {title} {Problem of the
  disordered chain},}\ }\href {\doibase 10.1063/1.1665300} {\bibfield
  {journal} {\bibinfo  {journal} {J. Math. Phys.}\ }\textbf {\bibinfo {volume}
  {11}},\ \bibinfo {pages} {1584--1590} (\bibinfo {year} {1970})}\BibitemShut
  {NoStop}%
\bibitem [{\citenamefont {Edwards}\ and\ \citenamefont
  {Jones}(1976)}]{edwards1976}%
  \BibitemOpen
  \bibfield  {author} {\bibinfo {author} {\bibfnamefont {S.~F.}\ \bibnamefont
  {Edwards}}\ and\ \bibinfo {author} {\bibfnamefont {R.~C.}\ \bibnamefont
  {Jones}},\ }\bibfield  {title} {\enquote {\bibinfo {title} {The eigenvalue
  spectrum of a large symmetric random matrix},}\ }\href@noop {} {\bibfield
  {journal} {\bibinfo  {journal} {J. Phys. A}\ }\textbf {\bibinfo {volume}
  {9}},\ \bibinfo {pages} {1595} (\bibinfo {year} {1976})}\BibitemShut
  {NoStop}%
\bibitem [{\citenamefont {Livan}\ \emph {et~al.}(2018)\citenamefont {Livan},
  \citenamefont {Novaes},\ and\ \citenamefont {Vivo}}]{livan2018}%
  \BibitemOpen
  \bibfield  {author} {\bibinfo {author} {\bibfnamefont {G.}~\bibnamefont
  {Livan}}, \bibinfo {author} {\bibfnamefont {M.}~\bibnamefont {Novaes}}, \
  and\ \bibinfo {author} {\bibfnamefont {P.}~\bibnamefont {Vivo}},\ }\href@noop
  {} {\emph {\bibinfo {title} {Introduction to Random Matrices: Theory and
  Practice}}}\ (\bibinfo  {publisher} {Springer International Publishing},\
  \bibinfo {year} {2018})\BibitemShut {NoStop}%
\bibitem [{\citenamefont {Potters}\ and\ \citenamefont
  {Bouchaud}(2020)}]{bouchaud2020}%
  \BibitemOpen
  \bibfield  {author} {\bibinfo {author} {\bibfnamefont {M.}~\bibnamefont
  {Potters}}\ and\ \bibinfo {author} {\bibfnamefont {J.-P.}\ \bibnamefont
  {Bouchaud}},\ }\href@noop {} {\emph {\bibinfo {title} {A First Course in
  Random Matrix Theory: for Physicists, Engineers and Data Scientists}}}\
  (\bibinfo  {publisher} {Cambridge University Press},\ \bibinfo {year}
  {2020})\BibitemShut {NoStop}%
\bibitem [{\citenamefont {Allen}(1999)}]{allen1999}%
  \BibitemOpen
  \bibfield  {author} {\bibinfo {author} {\bibfnamefont {Geoffrey}\
  \bibnamefont {Allen}},\ }\bibfield  {title} {\enquote {\bibinfo {title}
  {Geoffrey {G}ee, {C. B. E.} 6 {J}une 1910-13 {D}ecember 1996},}\ }\href@noop
  {} {\bibfield  {journal} {\bibinfo  {journal} {Biogr. mem. Fellows R. Soc.}\
  }\textbf {\bibinfo {volume} {45}},\ \bibinfo {pages} {185--194} (\bibinfo
  {year} {1999})}\BibitemShut {NoStop}%
\bibitem [{\citenamefont {Goldbart}\ and\ \citenamefont
  {Goldenfeld}(2004)}]{goldbart2004}%
  \BibitemOpen
  \bibfield  {author} {\bibinfo {author} {\bibfnamefont {P.~M.}\ \bibnamefont
  {Goldbart}}\ and\ \bibinfo {author} {\bibfnamefont {N.}~\bibnamefont
  {Goldenfeld}},\ }\bibfield  {title} {\enquote {\bibinfo {title} {{Sam Edwards
  and the Statistical Mechanics of Rubber}},}\ }in\ \href {\doibase
  10.1093/acprof:oso/9780198528531.003.0019} {\emph {\bibinfo {booktitle}
  {{Stealing the Gold: A celebration of the pioneering physics of Sam
  Edwards}}}},\ \bibinfo {editor} {edited by\ \bibinfo {editor} {\bibfnamefont
  {P.~M.}\ \bibnamefont {Goldbart}}, \bibinfo {editor} {\bibfnamefont
  {N.}~\bibnamefont {Goldenfeld}}, \ and\ \bibinfo {editor} {\bibfnamefont
  {D.}~\bibnamefont {Sherrington}}}\ (\bibinfo  {publisher} {Oxford University
  Press},\ \bibinfo {year} {2004})\BibitemShut {NoStop}%
\bibitem [{\citenamefont {Edwards}(1971)}]{edwards1971}%
  \BibitemOpen
  \bibfield  {author} {\bibinfo {author} {\bibfnamefont {S.~F.}\ \bibnamefont
  {Edwards}},\ }\bibfield  {title} {\enquote {\bibinfo {title} {The statistical
  mechanics of rubbers},}\ }in\ \href@noop {} {\emph {\bibinfo {booktitle}
  {Polymer Networks: Structure and Mechanical Properties-- Proceedings of the
  ACS Symposium on Highly Cross-Linked Polymer Networks, held in Chicago,
  Illinois, September 14–15, 1970}}},\ \bibinfo {editor} {edited by\ \bibinfo
  {editor} {\bibfnamefont {A.~J.}\ \bibnamefont {Chomp}}\ and\ \bibinfo
  {editor} {\bibfnamefont {S.}~\bibnamefont {Newman}}}\ (\bibinfo  {publisher}
  {Plenum Press},\ \bibinfo {year} {1971})\ pp.\ \bibinfo {pages}
  {83--110}\BibitemShut {NoStop}%
\bibitem [{\citenamefont {Edwards}(1972)}]{edwards1972}%
  \BibitemOpen
  \bibfield  {author} {\bibinfo {author} {\bibfnamefont {S.~F.}\ \bibnamefont
  {Edwards}},\ }\bibfield  {title} {\enquote {\bibinfo {title} {Statistical
  mechanics of polymerized materials},}\ }in\ \href@noop {} {\emph {\bibinfo
  {booktitle} {Amorphous materials: papers presented to the Third International
  Conference on the Physics of Non-crystalline Solids held at Sheffield
  University, September, 1970}}},\ \bibinfo {editor} {edited by\ \bibinfo
  {editor} {\bibfnamefont {R.~W.}\ \bibnamefont {Douglas}}\ and\ \bibinfo
  {editor} {\bibfnamefont {B.}~\bibnamefont {Ellis}}}\ (\bibinfo  {publisher}
  {Wiley-Interscience},\ \bibinfo {year} {1972})\ pp.\ \bibinfo {pages}
  {279--300}\BibitemShut {NoStop}%
\bibitem [{\citenamefont {Deam}(1975)}]{Deam1975}%
  \BibitemOpen
  \bibfield  {author} {\bibinfo {author} {\bibfnamefont {Rowan~Thomas}\
  \bibnamefont {Deam}},\ }\emph {\bibinfo {title} {Entanglements and excluded
  volume in rubber}},\ \href@noop {} {Ph.D. thesis},\ \bibinfo {address}
  {University of Cambridge} (\bibinfo {year} {1975})\BibitemShut {NoStop}%
\bibitem [{\citenamefont {Warner}(2017)}]{warner2017}%
  \BibitemOpen
  \bibfield  {author} {\bibinfo {author} {\bibfnamefont {M.}~\bibnamefont
  {Warner}},\ }\bibfield  {title} {\enquote {\bibinfo {title} {Sir {S}am
  {E}dwards. 1 {F}ebruary 1928--7 {J}uly 2015},}\ }\href {\doibase
  10.1098/rsbm.2016.0028} {\bibfield  {journal} {\bibinfo  {journal} {Biogr.
  mem. Fellows R. Soc.}\ }\textbf {\bibinfo {volume} {63}},\ \bibinfo {pages}
  {243--271} (\bibinfo {year} {2017})}\BibitemShut {NoStop}%
\bibitem [{\citenamefont {Deam}\ and\ \citenamefont
  {Edwards}(1976)}]{deam1976}%
  \BibitemOpen
  \bibfield  {author} {\bibinfo {author} {\bibfnamefont {R.~T.}\ \bibnamefont
  {Deam}}\ and\ \bibinfo {author} {\bibfnamefont {S.~F.}\ \bibnamefont
  {Edwards}},\ }\bibfield  {title} {\enquote {\bibinfo {title} {The theory of
  rubber elasticity},}\ }\href {\doibase 10.1098/rsta.1976.0001} {\bibfield
  {journal} {\bibinfo  {journal} {Philos. Trans. Royal Soc. A}\ }\textbf
  {\bibinfo {volume} {280}},\ \bibinfo {pages} {317--353} (\bibinfo {year}
  {1976})}\BibitemShut {NoStop}%
\bibitem [{\citenamefont {Ma}(1972)}]{ma1972}%
  \BibitemOpen
  \bibfield  {author} {\bibinfo {author} {\bibfnamefont {Shang-keng}\
  \bibnamefont {Ma}},\ }\bibfield  {title} {\enquote {\bibinfo {title}
  {Critical exponents for charged and neutral {B}ose gases above
  $\ensuremath{\lambda}$ points},}\ }\href {\doibase
  10.1103/PhysRevLett.29.1311} {\bibfield  {journal} {\bibinfo  {journal}
  {Phys. Rev. Lett.}\ }\textbf {\bibinfo {volume} {29}},\ \bibinfo {pages}
  {1311--1314} (\bibinfo {year} {1972})}\BibitemShut {NoStop}%
\bibitem [{\citenamefont {Grinstein}(1974)}]{grinstein1974}%
  \BibitemOpen
  \bibfield  {author} {\bibinfo {author} {\bibfnamefont {Geoffrey~Mark}\
  \bibnamefont {Grinstein}},\ }\emph {\bibinfo {title} {Magnetic phase
  transitions in alloys: A renormalization group approach}},\ \href@noop {}
  {Ph.D. thesis},\ \bibinfo {address} {Harvard University} (\bibinfo {year}
  {1974})\BibitemShut {NoStop}%
\bibitem [{\citenamefont {Ferrell}\ and\ \citenamefont
  {Scalapino}(1972)}]{ferrell1972}%
  \BibitemOpen
  \bibfield  {author} {\bibinfo {author} {\bibfnamefont {R.~A.}\ \bibnamefont
  {Ferrell}}\ and\ \bibinfo {author} {\bibfnamefont {D.~J.}\ \bibnamefont
  {Scalapino}},\ }\bibfield  {title} {\enquote {\bibinfo {title} {Screening
  solution in the field theory of phase transitions},}\ }\href {\doibase
  10.1016/0375-9601(72)90934-6} {\bibfield  {journal} {\bibinfo  {journal}
  {Phys. Lett. A}\ }\textbf {\bibinfo {volume} {41}},\ \bibinfo {pages}
  {371--372} (\bibinfo {year} {1972})}\BibitemShut {NoStop}%
\bibitem [{\citenamefont {Rivier}\ and\ \citenamefont
  {Teixeira}(1977)}]{rivier1977}%
  \BibitemOpen
  \bibfield  {author} {\bibinfo {author} {\bibfnamefont {N.}~\bibnamefont
  {Rivier}}\ and\ \bibinfo {author} {\bibfnamefont {N.~L.}\ \bibnamefont
  {Teixeira}},\ }\bibfield  {title} {\enquote {\bibinfo {title} {Electron
  localization in disordered one-dimensional systems and solitary waves in
  {G}inzburg-{L}andau ($\phi^4$) field theory},}\ }\href {\doibase
  10.1016/0375-9601(77)90944-6} {\bibfield  {journal} {\bibinfo  {journal}
  {Phys. Lett. A}\ }\textbf {\bibinfo {volume} {63}},\ \bibinfo {pages}
  {395--397} (\bibinfo {year} {1977})}\BibitemShut {NoStop}%
\bibitem [{\citenamefont {Ma}(1976)}]{ma1976}%
  \BibitemOpen
  \bibfield  {author} {\bibinfo {author} {\bibfnamefont {S.}~\bibnamefont
  {Ma}},\ }\href@noop {} {\emph {\bibinfo {title} {Modern Theory of Critical
  Phenomena}}}\ (\bibinfo  {publisher} {W. A. Benjamin},\ \bibinfo {year}
  {1976})\BibitemShut {NoStop}%
\bibitem [{\citenamefont {de~Gennes}(1972)}]{degennes1972}%
  \BibitemOpen
  \bibfield  {author} {\bibinfo {author} {\bibfnamefont {P.-G.}\ \bibnamefont
  {de~Gennes}},\ }\bibfield  {title} {\enquote {\bibinfo {title} {Exponents for
  the excluded volume problem as derived by the wilson method},}\ }\href@noop
  {} {\bibfield  {journal} {\bibinfo  {journal} {Phys. Lett. A}\ }\textbf
  {\bibinfo {volume} {38}},\ \bibinfo {pages} {339--340} (\bibinfo {year}
  {1972})}\BibitemShut {NoStop}%
\bibitem [{\citenamefont {Emery}(1975)}]{emery1975}%
  \BibitemOpen
  \bibfield  {author} {\bibinfo {author} {\bibfnamefont {V.~J.}\ \bibnamefont
  {Emery}},\ }\bibfield  {title} {\enquote {\bibinfo {title} {Critical
  properties of many-component systems},}\ }\href@noop {} {\bibfield  {journal}
  {\bibinfo  {journal} {Phys. Rev. B}\ }\textbf {\bibinfo {volume} {11}},\
  \bibinfo {pages} {239} (\bibinfo {year} {1975})}\BibitemShut {NoStop}%
\bibitem [{\citenamefont {Grinstein}\ and\ \citenamefont
  {Luther}(1976)}]{grinstein1976}%
  \BibitemOpen
  \bibfield  {author} {\bibinfo {author} {\bibfnamefont {G.}~\bibnamefont
  {Grinstein}}\ and\ \bibinfo {author} {\bibfnamefont {A.}~\bibnamefont
  {Luther}},\ }\bibfield  {title} {\enquote {\bibinfo {title} {Application of
  the renormalization group to phase transitions in disordered systems},}\
  }\href {\doibase 10.1103/PhysRevB.13.1329} {\bibfield  {journal} {\bibinfo
  {journal} {Phys. Rev. B}\ }\textbf {\bibinfo {volume} {13}},\ \bibinfo
  {pages} {1329--1343} (\bibinfo {year} {1976})}\BibitemShut {NoStop}%
\bibitem [{\citenamefont {Anderson}(1970)}]{anderson1970}%
  \BibitemOpen
  \bibfield  {author} {\bibinfo {author} {\bibfnamefont {P.~W.}\ \bibnamefont
  {Anderson}},\ }\bibfield  {title} {\enquote {\bibinfo {title} {Localisation
  theory and the {C}u-{M}n problem: Spin glasses},}\ }\href {\doibase
  10.1016/0025-5408(70)90096-6} {\bibfield  {journal} {\bibinfo  {journal}
  {Materials Research Bulletin}\ }\textbf {\bibinfo {volume} {5}},\ \bibinfo
  {pages} {549--554} (\bibinfo {year} {1970})}\BibitemShut {NoStop}%
\bibitem [{\citenamefont {Anderson}(1973)}]{anderson1972}%
  \BibitemOpen
  \bibfield  {author} {\bibinfo {author} {\bibfnamefont {P.~W.}\ \bibnamefont
  {Anderson}},\ }\bibfield  {title} {\enquote {\bibinfo {title} {Topics in spin
  glasses},}\ }in\ \href@noop {} {\emph {\bibinfo {booktitle} {Amorphous
  magnetism: Proceedings of the International Symposium on Amorphous Magnetism,
  August 17­18, 1972, Detroit, Michigan}}}\ (\bibinfo  {publisher} {Plenum
  Press},\ \bibinfo {year} {1973})\ pp.\ \bibinfo {pages} {1--14}\BibitemShut
  {NoStop}%
\bibitem [{\citenamefont {Anderson}\ \emph {et~al.}(1972)\citenamefont
  {Anderson}, \citenamefont {Halperin},\ and\ \citenamefont
  {Varma}}]{anderson1972anomalous}%
  \BibitemOpen
  \bibfield  {author} {\bibinfo {author} {\bibfnamefont {P.~W.}\ \bibnamefont
  {Anderson}}, \bibinfo {author} {\bibfnamefont {B.~I.}\ \bibnamefont
  {Halperin}}, \ and\ \bibinfo {author} {\bibfnamefont {C.~M.}\ \bibnamefont
  {Varma}},\ }\bibfield  {title} {\enquote {\bibinfo {title} {Anomalous
  low-temperature thermal properties of glasses and spin glasses},}\
  }\href@noop {} {\bibfield  {journal} {\bibinfo  {journal} {Philo. Mag.}\
  }\textbf {\bibinfo {volume} {25}},\ \bibinfo {pages} {1--9} (\bibinfo {year}
  {1972})}\BibitemShut {NoStop}%
\bibitem [{\citenamefont {Anderson}(2004)}]{anderson2004}%
  \BibitemOpen
  \bibfield  {author} {\bibinfo {author} {\bibfnamefont {P.~W.}\ \bibnamefont
  {Anderson}},\ }\bibfield  {title} {\enquote {\bibinfo {title} {{Remarks on
  the Edwards-Anderson Paper}},}\ }in\ \href {\doibase
  10.1093/acprof:oso/9780198528531.003.0014} {\emph {\bibinfo {booktitle}
  {{Stealing the Gold: A celebration of the pioneering physics of Sam
  Edwards}}}}\ (\bibinfo  {publisher} {Oxford University Press},\ \bibinfo
  {year} {2004})\BibitemShut {NoStop}%
\bibitem [{\citenamefont {Cannella}\ and\ \citenamefont
  {Mydosh}(1972)}]{cannella1972}%
  \BibitemOpen
  \bibfield  {author} {\bibinfo {author} {\bibfnamefont {V.}~\bibnamefont
  {Cannella}}\ and\ \bibinfo {author} {\bibfnamefont {J.~A.}\ \bibnamefont
  {Mydosh}},\ }\bibfield  {title} {\enquote {\bibinfo {title} {Magnetic
  ordering in gold-iron alloys},}\ }\href@noop {} {\bibfield  {journal}
  {\bibinfo  {journal} {Phys. Rev. B}\ }\textbf {\bibinfo {volume} {6}},\
  \bibinfo {pages} {4220} (\bibinfo {year} {1972})}\BibitemShut {NoStop}%
\bibitem [{\citenamefont {Charbonneau}(2021{\natexlab{a}})}]{mydosh2021}%
  \BibitemOpen
  \bibfield  {author} {\bibinfo {author} {\bibfnamefont {P.}~\bibnamefont
  {Charbonneau}},\ }\bibfield  {title} {\enquote {\bibinfo {title} {History of
  {RSB} interview: {L}eo van {H}emmen},}\ }\href {\doibase
  10.34847/nkl.e1e3ob87} {\bibfield  {journal} {\bibinfo  {journal} {History of
  RSB Project, CAPH\'ES, \'Ecole normale sup\'erieure, Paris}\ }\textbf
  {\bibinfo {volume} {19 p.}} (\bibinfo {year} {2021}{\natexlab{a}}),\
  10.34847/nkl.e1e3ob87},\ \bibinfo {note} {{T}ranscript of an oral history
  conducted 2021 by Patrick Charbonneau}\BibitemShut {NoStop}%
\bibitem [{\citenamefont {Edwards}\ and\ \citenamefont
  {Anderson}(1975)}]{edwards1975}%
  \BibitemOpen
  \bibfield  {author} {\bibinfo {author} {\bibfnamefont {S.~F.}\ \bibnamefont
  {Edwards}}\ and\ \bibinfo {author} {\bibfnamefont {P.~W.}\ \bibnamefont
  {Anderson}},\ }\bibfield  {title} {\enquote {\bibinfo {title} {Theory of spin
  glasses},}\ }\href@noop {} {\bibfield  {journal} {\bibinfo  {journal} {J.
  Phys. F}\ }\textbf {\bibinfo {volume} {5}},\ \bibinfo {pages} {965} (\bibinfo
  {year} {1975})}\BibitemShut {NoStop}%
\bibitem [{\citenamefont {Zangwill}(2021)}]{zangwill2021}%
  \BibitemOpen
  \bibfield  {author} {\bibinfo {author} {\bibfnamefont {A.}~\bibnamefont
  {Zangwill}},\ }\href@noop {} {\emph {\bibinfo {title} {A Mind over matter}}}\
  (\bibinfo  {publisher} {Oxford University Press},\ \bibinfo {year}
  {2021})\BibitemShut {NoStop}%
\bibitem [{\citenamefont {Charbonneau}(2021{\natexlab{b}})}]{sherrington2020}%
  \BibitemOpen
  \bibfield  {author} {\bibinfo {author} {\bibfnamefont {P.}~\bibnamefont
  {Charbonneau}},\ }\bibfield  {title} {\enquote {\bibinfo {title} {History of
  {RSB} interview: {D}avid {S}herrington},}\ }\href {\doibase
  10.34847/nkl.072dc5a6} {\bibfield  {journal} {\bibinfo  {journal} {History of
  RSB Project, CAPH\'ES, \'Ecole normale sup\'erieure, Paris}\ }\textbf
  {\bibinfo {volume} {39 p.}} (\bibinfo {year} {2021}{\natexlab{b}}),\
  10.34847/nkl.072dc5a6},\ \bibinfo {note} {{T}ranscript of an oral history
  conducted 2020 by Patrick Charbonneau and Francesco Zamponi}\BibitemShut
  {NoStop}%
\bibitem [{\citenamefont {Adkins}(1974)}]{adkins1974thesis}%
  \BibitemOpen
  \bibfield  {author} {\bibinfo {author} {\bibfnamefont {Keith~John}\
  \bibnamefont {Adkins}},\ }\emph {\bibinfo {title} {Theory of spin glasses}},\
  \href@noop {} {Ph.D. thesis},\ \bibinfo  {school} {Imperial College London}
  (\bibinfo {year} {1974})\BibitemShut {NoStop}%
\bibitem [{\citenamefont {Adkins}\ and\ \citenamefont
  {Rivier}(1974)}]{adkins1974}%
  \BibitemOpen
  \bibfield  {author} {\bibinfo {author} {\bibfnamefont {K.}~\bibnamefont
  {Adkins}}\ and\ \bibinfo {author} {\bibfnamefont {N.}~\bibnamefont
  {Rivier}},\ }\bibfield  {title} {\enquote {\bibinfo {title} {Susceptibility
  of spn glasses},}\ }\href {\doibase 10.1051/jphyscol:1974443} {\bibfield
  {journal} {\bibinfo  {journal} {J. Phys. Colloques}\ }\textbf {\bibinfo
  {volume} {35}},\ \bibinfo {pages} {C4--237--C4--240} (\bibinfo {year}
  {1974})}\BibitemShut {NoStop}%
\bibitem [{\citenamefont {Warner}(1977)}]{warner1977}%
  \BibitemOpen
  \bibfield  {author} {\bibinfo {author} {\bibfnamefont {Mark}\ \bibnamefont
  {Warner}},\ }\emph {\bibinfo {title} {Molecular motion of polymeric
  systems}},\ \href {http://hdl.handle.net/10044/1/22821} {Ph.D. thesis},\
  \bibinfo  {school} {Imperial College London} (\bibinfo {year}
  {1977})\BibitemShut {NoStop}%
\bibitem [{\citenamefont {Sherrington}\ and\ \citenamefont
  {Mihill}(1974)}]{sherrington1974}%
  \BibitemOpen
  \bibfield  {author} {\bibinfo {author} {\bibfnamefont {D.}~\bibnamefont
  {Sherrington}}\ and\ \bibinfo {author} {\bibfnamefont {K.}~\bibnamefont
  {Mihill}},\ }\bibfield  {title} {\enquote {\bibinfo {title} {Effects of
  clustering on the magnetic properties of transition metal alloys},}\ }\href
  {\doibase 10.1051/jphyscol:1974435} {\bibfield  {journal} {\bibinfo
  {journal} {J. Phys. Colloques}\ }\textbf {\bibinfo {volume} {35}},\ \bibinfo
  {pages} {C4--199--C4--201} (\bibinfo {year} {1974})}\BibitemShut {NoStop}%
\bibitem [{\citenamefont {Caplin}(1999)}]{coles1999}%
  \BibitemOpen
  \bibfield  {author} {\bibinfo {author} {\bibfnamefont {D.}~\bibnamefont
  {Caplin}},\ }\bibfield  {title} {\enquote {\bibinfo {title} {Bryan {R}andell
  {C}oles. 9 {J}une 1926-24 {F}ebruary 1997},}\ }\href@noop {} {\bibfield
  {journal} {\bibinfo  {journal} {Biogr. mem. Fellows R. Soc.}\ }\textbf
  {\bibinfo {volume} {45}},\ \bibinfo {pages} {53--66} (\bibinfo {year}
  {1999})}\BibitemShut {NoStop}%
\bibitem [{\citenamefont {Anderson}(1992)}]{anderson1992}%
  \BibitemOpen
  \bibfield  {author} {\bibinfo {author} {\bibfnamefont {P.~W.}\ \bibnamefont
  {Anderson}},\ }\bibfield  {title} {\enquote {\bibinfo {title}
  {Introduction},}\ }in\ \href@noop {} {\emph {\bibinfo {booktitle} {Spin
  Glasses and Biology}}}\ (\bibinfo  {publisher} {World Scientific
  Publishing},\ \bibinfo {year} {1992})\ pp.\ \bibinfo {pages}
  {1--5}\BibitemShut {NoStop}%
\bibitem [{\citenamefont {Sherrington}(1975)}]{sherrington1975b}%
  \BibitemOpen
  \bibfield  {author} {\bibinfo {author} {\bibfnamefont {D.}~\bibnamefont
  {Sherrington}},\ }\bibfield  {title} {\enquote {\bibinfo {title} {A
  transparent theory of the spin glass},}\ }\href {\doibase
  10.1088/0022-3719/8/10/021} {\bibfield  {journal} {\bibinfo  {journal} {J.
  Phys. C}\ }\textbf {\bibinfo {volume} {8}},\ \bibinfo {pages} {L208--L212}
  (\bibinfo {year} {1975})}\BibitemShut {NoStop}%
\bibitem [{\citenamefont {Sherrington}\ and\ \citenamefont
  {Southern}(1975)}]{sherrington1975c}%
  \BibitemOpen
  \bibfield  {author} {\bibinfo {author} {\bibfnamefont {D.}~\bibnamefont
  {Sherrington}}\ and\ \bibinfo {author} {\bibfnamefont {B.~W.}\ \bibnamefont
  {Southern}},\ }\bibfield  {title} {\enquote {\bibinfo {title} {Spin glass
  versus ferromagnet},}\ }\href@noop {} {\bibfield  {journal} {\bibinfo
  {journal} {J. Phys. F}\ }\textbf {\bibinfo {volume} {5}},\ \bibinfo {pages}
  {L49} (\bibinfo {year} {1975})}\BibitemShut {NoStop}%
\bibitem [{\citenamefont {Charbonneau}(2021{\natexlab{c}})}]{southern2021}%
  \BibitemOpen
  \bibfield  {author} {\bibinfo {author} {\bibfnamefont {P.}~\bibnamefont
  {Charbonneau}},\ }\bibfield  {title} {\enquote {\bibinfo {title} {History of
  {RSB} interview: {B}yron {W}. {S}outhern},}\ }\href {\doibase
  10.34847/nkl.1f8a00ei} {\bibfield  {journal} {\bibinfo  {journal} {History of
  RSB Project, CAPH\'ES, \'Ecole normale sup\'erieure, Paris}\ }\textbf
  {\bibinfo {volume} {9 p.}} (\bibinfo {year} {2021}{\natexlab{c}}),\
  10.34847/nkl.1f8a00ei},\ \bibinfo {note} {{T}ranscript of an oral history
  conducted 2021 by Patrick Charbonneau and Francesco Zamponi}\BibitemShut
  {NoStop}%
\bibitem [{\citenamefont {Charbonneau}(2021{\natexlab{d}})}]{kirkpatrick2021}%
  \BibitemOpen
  \bibfield  {author} {\bibinfo {author} {\bibfnamefont {P.}~\bibnamefont
  {Charbonneau}},\ }\bibfield  {title} {\enquote {\bibinfo {title} {History of
  {RSB} interview: {S}cott {K}irkpatrick},}\ }\href {\doibase
  10.34847/nkl.cba615t7} {\bibfield  {journal} {\bibinfo  {journal} {History of
  RSB Project, CAPH\'ES, \'Ecole normale sup\'erieure, Paris}\ }\textbf
  {\bibinfo {volume} {24 p.}} (\bibinfo {year} {2021}{\natexlab{d}}),\
  10.34847/nkl.cba615t7},\ \bibinfo {note} {{T}ranscript of an oral history
  conducted 2021 by Patrick Charbonneau and Francesco Zamponi}\BibitemShut
  {NoStop}%
\bibitem [{\citenamefont {Sherrington}\ and\ \citenamefont
  {Kirkpatrick}(1975)}]{sherrington1975}%
  \BibitemOpen
  \bibfield  {author} {\bibinfo {author} {\bibfnamefont {D.}~\bibnamefont
  {Sherrington}}\ and\ \bibinfo {author} {\bibfnamefont {S.}~\bibnamefont
  {Kirkpatrick}},\ }\bibfield  {title} {\enquote {\bibinfo {title} {Solvable
  model of a spin-glass},}\ }\href@noop {} {\bibfield  {journal} {\bibinfo
  {journal} {Phys. Rev. Lett.}\ }\textbf {\bibinfo {volume} {35}},\ \bibinfo
  {pages} {1792--1796} (\bibinfo {year} {1975})}\BibitemShut {NoStop}%
\bibitem [{\citenamefont {de~Almeida}\ and\ \citenamefont
  {Thouless}(1978)}]{dealmeida1978}%
  \BibitemOpen
  \bibfield  {author} {\bibinfo {author} {\bibfnamefont {J.~R.~L.}\
  \bibnamefont {de~Almeida}}\ and\ \bibinfo {author} {\bibfnamefont {D.~J.}\
  \bibnamefont {Thouless}},\ }\bibfield  {title} {\enquote {\bibinfo {title}
  {Stability of the sherrington-kirkpatrick solution of a spin glass model},}\
  }\href@noop {} {\bibfield  {journal} {\bibinfo  {journal} {J. Phys. A}\
  }\textbf {\bibinfo {volume} {11}},\ \bibinfo {pages} {983} (\bibinfo {year}
  {1978})}\BibitemShut {NoStop}%
\bibitem [{\citenamefont {Charbonneau}(2021{\natexlab{e}})}]{dealmeida2021}%
  \BibitemOpen
  \bibfield  {author} {\bibinfo {author} {\bibfnamefont {P.}~\bibnamefont
  {Charbonneau}},\ }\bibfield  {title} {\enquote {\bibinfo {title} {History of
  {RSB} interview: {J}airo de {A}lmeida},}\ }\href {\doibase
  10.34847/nkl.7de8emt7} {\bibfield  {journal} {\bibinfo  {journal} {History of
  RSB Project, CAPH\'ES, \'Ecole normale sup\'erieure, Paris}\ }\textbf
  {\bibinfo {volume} {23 p.}} (\bibinfo {year} {2021}{\natexlab{e}}),\
  10.34847/nkl.7de8emt7},\ \bibinfo {note} {{T}ranscript of an oral history
  conducted 2021 by Patrick Charbonneau and Francesco Zamponi}\BibitemShut
  {NoStop}%
\bibitem [{\citenamefont {Charbonneau}(2021{\natexlab{f}})}]{moore2020}%
  \BibitemOpen
  \bibfield  {author} {\bibinfo {author} {\bibfnamefont {P.}~\bibnamefont
  {Charbonneau}},\ }\bibfield  {title} {\enquote {\bibinfo {title} {History of
  {RSB} interview: {M}ichael {M}oore},}\ }\href {\doibase
  10.34847/nkl.997eiv27} {\bibfield  {journal} {\bibinfo  {journal} {History of
  RSB Project, CAPH\'ES, \'Ecole normale sup\'erieure, Paris}\ }\textbf
  {\bibinfo {volume} {26 p.}} (\bibinfo {year} {2021}{\natexlab{f}}),\
  10.34847/nkl.997eiv27},\ \bibinfo {note} {{T}ranscript of an oral history
  conducted 2020 by Patrick Charbonneau and Francesco Zamponi}\BibitemShut
  {NoStop}%
\bibitem [{\citenamefont {van Hemmen}\ and\ \citenamefont
  {Palmer}(1979)}]{vanhemmen1979}%
  \BibitemOpen
  \bibfield  {author} {\bibinfo {author} {\bibfnamefont {J.~L.}\ \bibnamefont
  {van Hemmen}}\ and\ \bibinfo {author} {\bibfnamefont {R.~G.}\ \bibnamefont
  {Palmer}},\ }\bibfield  {title} {\enquote {\bibinfo {title} {The replica
  method and solvable spin glass model},}\ }\href@noop {} {\bibfield  {journal}
  {\bibinfo  {journal} {J. Phys. A}\ }\textbf {\bibinfo {volume} {12}},\
  \bibinfo {pages} {563--580} (\bibinfo {year} {1979})}\BibitemShut {NoStop}%
\bibitem [{\citenamefont {Charbonneau}(2021{\natexlab{g}})}]{vanhemmen2021}%
  \BibitemOpen
  \bibfield  {author} {\bibinfo {author} {\bibfnamefont {P.}~\bibnamefont
  {Charbonneau}},\ }\bibfield  {title} {\enquote {\bibinfo {title} {History of
  {RSB} interview: {L}eo van {H}emmen},}\ }\href {\doibase
  10.34847/nkl.16e5m0oj} {\bibfield  {journal} {\bibinfo  {journal} {History of
  RSB Project, CAPH\'ES, \'Ecole normale sup\'erieure, Paris}\ }\textbf
  {\bibinfo {volume} {22 p.}} (\bibinfo {year} {2021}{\natexlab{g}}),\
  10.34847/nkl.16e5m0oj},\ \bibinfo {note} {{T}ranscript of an oral history
  conducted 2021 by Patrick Charbonneau and Francesco Zamponi}\BibitemShut
  {NoStop}%
\bibitem [{\citenamefont {Thouless}\ \emph {et~al.}(1977)\citenamefont
  {Thouless}, \citenamefont {Anderson},\ and\ \citenamefont
  {Palmer}}]{thouless1977}%
  \BibitemOpen
  \bibfield  {author} {\bibinfo {author} {\bibfnamefont {D.~J.}\ \bibnamefont
  {Thouless}}, \bibinfo {author} {\bibfnamefont {P.~W.}\ \bibnamefont
  {Anderson}}, \ and\ \bibinfo {author} {\bibfnamefont {R.~G.}\ \bibnamefont
  {Palmer}},\ }\bibfield  {title} {\enquote {\bibinfo {title} {Solution of
  `solvable model of a spin glass'},}\ }\href@noop {} {\bibfield  {journal}
  {\bibinfo  {journal} {Philo. Mag.}\ }\textbf {\bibinfo {volume} {35}},\
  \bibinfo {pages} {593--601} (\bibinfo {year} {1977})}\BibitemShut {NoStop}%
\bibitem [{\citenamefont {Thouless}(1986)}]{thouless1986}%
  \BibitemOpen
  \bibfield  {author} {\bibinfo {author} {\bibfnamefont {D.~J.}\ \bibnamefont
  {Thouless}},\ }\bibfield  {title} {\enquote {\bibinfo {title} {Spin-glass on
  a {B}ethe lattice},}\ }\href@noop {} {\bibfield  {journal} {\bibinfo
  {journal} {Phys. Rev. Lett.}\ }\textbf {\bibinfo {volume} {56}},\ \bibinfo
  {pages} {1082} (\bibinfo {year} {1986})}\BibitemShut {NoStop}%
\bibitem [{\citenamefont {Chayes}\ \emph {et~al.}(1986)\citenamefont {Chayes},
  \citenamefont {Chayes}, \citenamefont {Sethna},\ and\ \citenamefont
  {Thouless}}]{chayes1986}%
  \BibitemOpen
  \bibfield  {author} {\bibinfo {author} {\bibfnamefont {J.T.}\ \bibnamefont
  {Chayes}}, \bibinfo {author} {\bibfnamefont {L.}~\bibnamefont {Chayes}},
  \bibinfo {author} {\bibfnamefont {J.P.}\ \bibnamefont {Sethna}}, \ and\
  \bibinfo {author} {\bibfnamefont {D.~J.}\ \bibnamefont {Thouless}},\
  }\bibfield  {title} {\enquote {\bibinfo {title} {A mean field spin glass with
  short-range interactions},}\ }\href@noop {} {\bibfield  {journal} {\bibinfo
  {journal} {Commun.Math. Phys.}\ }\textbf {\bibinfo {volume} {106}},\ \bibinfo
  {pages} {41–89} (\bibinfo {year} {1986})}\BibitemShut {NoStop}%
\bibitem [{\citenamefont {Carlson}\ \emph {et~al.}(1988)\citenamefont
  {Carlson}, \citenamefont {Chayes}, \citenamefont {Chayes}, \citenamefont
  {Sethna},\ and\ \citenamefont {Thouless}}]{carlson1988}%
  \BibitemOpen
  \bibfield  {author} {\bibinfo {author} {\bibfnamefont {J.~M.}\ \bibnamefont
  {Carlson}}, \bibinfo {author} {\bibfnamefont {J.~T.}\ \bibnamefont {Chayes}},
  \bibinfo {author} {\bibfnamefont {L.}~\bibnamefont {Chayes}}, \bibinfo
  {author} {\bibfnamefont {J.~P.}\ \bibnamefont {Sethna}}, \ and\ \bibinfo
  {author} {\bibfnamefont {D.~J.}\ \bibnamefont {Thouless}},\ }\bibfield
  {title} {\enquote {\bibinfo {title} {Critical behavior of the {B}ethe lattice
  spin glass},}\ }\href@noop {} {\bibfield  {journal} {\bibinfo  {journal}
  {Europhys. Lett.}\ }\textbf {\bibinfo {volume} {5}},\ \bibinfo {pages} {355}
  (\bibinfo {year} {1988})}\BibitemShut {NoStop}%
\bibitem [{\citenamefont {Kosterlitz}\ \emph {et~al.}(1976)\citenamefont
  {Kosterlitz}, \citenamefont {Thouless},\ and\ \citenamefont
  {Jones}}]{kosterlitz1976}%
  \BibitemOpen
  \bibfield  {author} {\bibinfo {author} {\bibfnamefont {J.~M.}\ \bibnamefont
  {Kosterlitz}}, \bibinfo {author} {\bibfnamefont {D.~J.}\ \bibnamefont
  {Thouless}}, \ and\ \bibinfo {author} {\bibfnamefont {R.~C.}\ \bibnamefont
  {Jones}},\ }\bibfield  {title} {\enquote {\bibinfo {title} {Spherical model
  of a spin-glass},}\ }\href@noop {} {\bibfield  {journal} {\bibinfo  {journal}
  {Phys. Rev. Lett.}\ }\textbf {\bibinfo {volume} {36}},\ \bibinfo {pages}
  {1217} (\bibinfo {year} {1976})}\BibitemShut {NoStop}%
\bibitem [{\citenamefont {Bray}\ and\ \citenamefont {Moore}(1978)}]{bray1978}%
  \BibitemOpen
  \bibfield  {author} {\bibinfo {author} {\bibfnamefont {A.~J.}\ \bibnamefont
  {Bray}}\ and\ \bibinfo {author} {\bibfnamefont {M.~A.}\ \bibnamefont
  {Moore}},\ }\bibfield  {title} {\enquote {\bibinfo {title} {Replica-symmetry
  breaking in spin-glass theories},}\ }\href {\doibase
  10.1103/PhysRevLett.41.1068} {\bibfield  {journal} {\bibinfo  {journal}
  {Phys. Rev. Lett.}\ }\textbf {\bibinfo {volume} {41}},\ \bibinfo {pages}
  {1068--1072} (\bibinfo {year} {1978})}\BibitemShut {NoStop}%
\bibitem [{\citenamefont {Blandin}(1961)}]{blandin1961}%
  \BibitemOpen
  \bibfield  {author} {\bibinfo {author} {\bibfnamefont {Andr\'e}\ \bibnamefont
  {Blandin}},\ }\emph {\bibinfo {title} {Contribution à l'\'etude de la
  structure \'electronique des impuret\'es dans les m\'etaux}},\ \href@noop {}
  {Ph.D. thesis},\ \bibinfo  {school} {Universit\'e de Paris} (\bibinfo {year}
  {1961})\BibitemShut {NoStop}%
\bibitem [{\citenamefont {Friedel}(1994)}]{friedel1994}%
  \BibitemOpen
  \bibfield  {author} {\bibinfo {author} {\bibfnamefont {J.}~\bibnamefont
  {Friedel}},\ }\href@noop {} {\emph {\bibinfo {title} {Graine de mandarin}}}\
  (\bibinfo  {publisher} {Editions Odile Jacob},\ \bibinfo {year}
  {1994})\BibitemShut {NoStop}%
\bibitem [{\citenamefont {Marshall}(1960)}]{marshall1960}%
  \BibitemOpen
  \bibfield  {author} {\bibinfo {author} {\bibfnamefont {W.}~\bibnamefont
  {Marshall}},\ }\bibfield  {title} {\enquote {\bibinfo {title} {Specific heat
  of dilute alloys},}\ }\href {\doibase 10.1103/PhysRev.118.1519} {\bibfield
  {journal} {\bibinfo  {journal} {Phys. Rev.}\ }\textbf {\bibinfo {volume}
  {118}},\ \bibinfo {pages} {1519--1523} (\bibinfo {year} {1960})}\BibitemShut
  {NoStop}%
\bibitem [{\citenamefont {Blandin}\ and\ \citenamefont
  {Friedel}(1959)}]{blandin1959}%
  \BibitemOpen
  \bibfield  {author} {\bibinfo {author} {\bibfnamefont {A.}~\bibnamefont
  {Blandin}}\ and\ \bibinfo {author} {\bibfnamefont {J.}~\bibnamefont
  {Friedel}},\ }\bibfield  {title} {\enquote {\bibinfo {title}
  {Propri{\'e}t{\'e}s magn{\'e}tiques des alliages dilu{\'e}s. interactions
  magn{\'e}tiques et antiferromagn{\'e}tisme dans les alliages du type
  m{\'e}tal noble-m{\'e}tal de transition},}\ }\href@noop {} {\bibfield
  {journal} {\bibinfo  {journal} {J. Phys. Radium}\ }\textbf {\bibinfo {volume}
  {20}},\ \bibinfo {pages} {160--168} (\bibinfo {year} {1959})}\BibitemShut
  {NoStop}%
\bibitem [{\citenamefont {Friedel}(1985)}]{friedel1985}%
  \BibitemOpen
  \bibfield  {author} {\bibinfo {author} {\bibfnamefont {J.}~\bibnamefont
  {Friedel}},\ }\bibfield  {title} {\enquote {\bibinfo {title} {André
  {B}landin 1933-1985},}\ }\href@noop {} {\bibfield  {journal} {\bibinfo
  {journal} {Ann. Phys. Fr.}\ }\textbf {\bibinfo {volume} {10}},\ \bibinfo
  {pages} {1--4} (\bibinfo {year} {1985})}\BibitemShut {NoStop}%
\bibitem [{\citenamefont {Blandin}(1978)}]{blandin1978}%
  \BibitemOpen
  \bibfield  {author} {\bibinfo {author} {\bibfnamefont {A.}~\bibnamefont
  {Blandin}},\ }\bibfield  {title} {\enquote {\bibinfo {title} {Theories versus
  experiments in the spin glass systems},}\ }\href {\doibase
  10.1051/jphyscol:19786593} {\bibfield  {journal} {\bibinfo  {journal} {J.
  Phys. Colloques}\ }\textbf {\bibinfo {volume} {39}},\ \bibinfo {pages}
  {C6--1499--C6--1516} (\bibinfo {year} {1978})}\BibitemShut {NoStop}%
\bibitem [{\citenamefont {Toulouse}(1985)}]{toulouse1985}%
  \BibitemOpen
  \bibfield  {author} {\bibinfo {author} {\bibfnamefont {G.}~\bibnamefont
  {Toulouse}},\ }\bibfield  {title} {\enquote {\bibinfo {title} {André
  {B}landin et la physique des verres de spin. {T}rois étés alpins : 1958,
  1968, 1978},}\ }\href@noop {} {\bibfield  {journal} {\bibinfo  {journal}
  {Ann. Phys. Fr.}\ }\textbf {\bibinfo {volume} {10}},\ \bibinfo {pages}
  {85--100} (\bibinfo {year} {1985})}\BibitemShut {NoStop}%
\bibitem [{\citenamefont {Charbonneau}(2021{\natexlab{h}})}]{gabay2021}%
  \BibitemOpen
  \bibfield  {author} {\bibinfo {author} {\bibfnamefont {P.}~\bibnamefont
  {Charbonneau}},\ }\bibfield  {title} {\enquote {\bibinfo {title} {History of
  {RSB} interview: {M}arc {G}abay},}\ }\href {\doibase 10.34847/nkl.f14cb3mt}
  {\bibfield  {journal} {\bibinfo  {journal} {History of RSB Project, CAPH\'ES,
  \'Ecole normale sup\'erieure, Paris}\ }\textbf {\bibinfo {volume} {18 p.}}
  (\bibinfo {year} {2021}{\natexlab{h}}),\ 10.34847/nkl.f14cb3mt},\ \bibinfo
  {note} {{T}ranscript of an oral history conducted 2021 by Patrick Charbonneau
  and Francesco Zamponi}\BibitemShut {NoStop}%
\bibitem [{\citenamefont {Blandin}\ \emph {et~al.}(1980)\citenamefont
  {Blandin}, \citenamefont {Gabay},\ and\ \citenamefont {Garel}}]{blandin1980}%
  \BibitemOpen
  \bibfield  {author} {\bibinfo {author} {\bibfnamefont {A.}~\bibnamefont
  {Blandin}}, \bibinfo {author} {\bibfnamefont {M.}~\bibnamefont {Gabay}}, \
  and\ \bibinfo {author} {\bibfnamefont {T.}~\bibnamefont {Garel}},\ }\bibfield
   {title} {\enquote {\bibinfo {title} {On the mean-field theory of spin
  glasses},}\ }\href@noop {} {\bibfield  {journal} {\bibinfo  {journal} {J.
  Phys. C}\ }\textbf {\bibinfo {volume} {13}},\ \bibinfo {pages} {403--418}
  (\bibinfo {year} {1980})}\BibitemShut {NoStop}%
\bibitem [{\citenamefont {Charbonneau}(2021{\natexlab{i}})}]{lubensky2021}%
  \BibitemOpen
  \bibfield  {author} {\bibinfo {author} {\bibfnamefont {P.}~\bibnamefont
  {Charbonneau}},\ }\bibfield  {title} {\enquote {\bibinfo {title} {History of
  {RSB} interview: {T}om {C}. {L}ubensky},}\ }\href {\doibase
  10.34847/nkl.f2cap2m9} {\bibfield  {journal} {\bibinfo  {journal} {History of
  RSB Project, CAPH\'ES, \'Ecole normale sup\'erieure, Paris}\ }\textbf
  {\bibinfo {volume} {13 p.}} (\bibinfo {year} {2021}{\natexlab{i}}),\
  10.34847/nkl.f2cap2m9},\ \bibinfo {note} {{T}ranscript of an oral history
  conducted 2021 by Patrick Charbonneau and Francesco Zamponi}\BibitemShut
  {NoStop}%
\bibitem [{\citenamefont {Kohn}(1990)}]{kohn1990}%
  \BibitemOpen
  \bibfield  {author} {\bibinfo {author} {\bibfnamefont {W.}~\bibnamefont
  {Kohn}},\ }\bibfield  {title} {\enquote {\bibinfo {title} {André, one of my
  dearest and closest friends},}\ }in\ \href@noop {} {\emph {\bibinfo
  {booktitle} {André Blandin 1933-1985}}}\ (\bibinfo  {publisher} {s.n.},\
  \bibinfo {year} {1990})\ pp.\ \bibinfo {pages} {18--20}\BibitemShut {NoStop}%
\bibitem [{\citenamefont {Kirkpatrick}\ and\ \citenamefont
  {Sherrington}(1978)}]{kirkpatrick1978}%
  \BibitemOpen
  \bibfield  {author} {\bibinfo {author} {\bibfnamefont {S.}~\bibnamefont
  {Kirkpatrick}}\ and\ \bibinfo {author} {\bibfnamefont {D.}~\bibnamefont
  {Sherrington}},\ }\bibfield  {title} {\enquote {\bibinfo {title}
  {Infinite-ranged models of spin-glasses},}\ }\href {\doibase
  10.1103/PhysRevB.17.4384} {\bibfield  {journal} {\bibinfo  {journal} {Phys.
  Rev. B}\ }\textbf {\bibinfo {volume} {17}},\ \bibinfo {pages} {4384--4403}
  (\bibinfo {year} {1978})}\BibitemShut {NoStop}%
\bibitem [{\citenamefont {Parisi}(1980{\natexlab{a}})}]{parisi1980}%
  \BibitemOpen
  \bibfield  {author} {\bibinfo {author} {\bibfnamefont {G.}~\bibnamefont
  {Parisi}},\ }\bibfield  {title} {\enquote {\bibinfo {title} {A sequence of
  approximated solutions to the {S}-{K} model for spin glasses},}\ }\href
  {\doibase 10.1088/0305-4470/13/4/009} {\bibfield  {journal} {\bibinfo
  {journal} {J. Phys. A}\ }\textbf {\bibinfo {volume} {13}},\ \bibinfo {pages}
  {L115--L121} (\bibinfo {year} {1980}{\natexlab{a}})}\BibitemShut {NoStop}%
\bibitem [{\citenamefont {Balian}\ \emph {et~al.}(1979)\citenamefont {Balian},
  \citenamefont {Maynard},\ and\ \citenamefont {Toulouse}}]{leshouches1978}%
  \BibitemOpen
  \bibinfo {editor} {\bibfnamefont {R.}~\bibnamefont {Balian}}, \bibinfo
  {editor} {\bibfnamefont {R.}~\bibnamefont {Maynard}}, \ and\ \bibinfo
  {editor} {\bibfnamefont {G.}~\bibnamefont {Toulouse}},\ eds.,\ \href@noop {}
  {\emph {\bibinfo {title} {La {M}atière mal condensée/{I}ll-{C}ondensed
  {M}atter}}}\ (\bibinfo  {publisher} {North-Holland Publishing Company},\
  \bibinfo {year} {1979})\BibitemShut {NoStop}%
\bibitem [{\citenamefont {Anderson}(1979)}]{anderson1978}%
  \BibitemOpen
  \bibfield  {author} {\bibinfo {author} {\bibfnamefont {P.~W.}\ \bibnamefont
  {Anderson}},\ }\bibfield  {title} {\enquote {\bibinfo {title} {Lectures on
  amorphous systems},}\ }in\ \href@noop {} {\emph {\bibinfo {booktitle} {La
  {M}atière mal condensée/{I}ll-{C}ondensed {M}atter}}},\ \bibinfo {editor}
  {edited by\ \bibinfo {editor} {\bibfnamefont {R.}~\bibnamefont {Balian}},
  \bibinfo {editor} {\bibfnamefont {R.}~\bibnamefont {Maynard}}, \ and\
  \bibinfo {editor} {\bibfnamefont {G.}~\bibnamefont {Toulouse}}}\ (\bibinfo
  {publisher} {North-Holland Publishing Company},\ \bibinfo {year} {1979})\
  pp.\ \bibinfo {pages} {159--262}\BibitemShut {NoStop}%
\bibitem [{\citenamefont {Charbonneau}(2021{\natexlab{j}})}]{derrida2020}%
  \BibitemOpen
  \bibfield  {author} {\bibinfo {author} {\bibfnamefont {P.}~\bibnamefont
  {Charbonneau}},\ }\bibfield  {title} {\enquote {\bibinfo {title} {History of
  {RSB} interview: {B}ernard {D}errida},}\ }\href {\doibase
  10.34847/nkl.3e183b0o} {\bibfield  {journal} {\bibinfo  {journal} {History of
  RSB Project, CAPH\'ES, \'Ecole normale sup\'erieure, Paris}\ }\textbf
  {\bibinfo {volume} {23 p.}} (\bibinfo {year} {2021}{\natexlab{j}}),\
  10.34847/nkl.3e183b0o},\ \bibinfo {note} {{T}ranscript of an oral history
  conducted 2020 by Patrick Charbonneau and Francesco Zamponi}\BibitemShut
  {NoStop}%
\bibitem [{\citenamefont {Chaudhari}\ \emph {et~al.}(1979)\citenamefont
  {Chaudhari}, \citenamefont {Levi},\ and\ \citenamefont
  {Steinhardt}}]{chaudhari1979}%
  \BibitemOpen
  \bibfield  {author} {\bibinfo {author} {\bibfnamefont {P.}~\bibnamefont
  {Chaudhari}}, \bibinfo {author} {\bibfnamefont {A.}~\bibnamefont {Levi}}, \
  and\ \bibinfo {author} {\bibfnamefont {P.}~\bibnamefont {Steinhardt}},\
  }\bibfield  {title} {\enquote {\bibinfo {title} {Edge and screw dislocations
  in an amorphous solid},}\ }\href {\doibase 10.1103/PhysRevLett.43.1517}
  {\bibfield  {journal} {\bibinfo  {journal} {Phys. Rev. Lett.}\ }\textbf
  {\bibinfo {volume} {43}},\ \bibinfo {pages} {1517--1520} (\bibinfo {year}
  {1979})}\BibitemShut {NoStop}%
\bibitem [{\citenamefont {d'Eramo}\ \emph {et~al.}(1971)\citenamefont
  {d'Eramo}, \citenamefont {Peliti},\ and\ \citenamefont {Parisi}}]{eramo1971}%
  \BibitemOpen
  \bibfield  {author} {\bibinfo {author} {\bibfnamefont {M.}~\bibnamefont
  {d'Eramo}}, \bibinfo {author} {\bibfnamefont {L.}~\bibnamefont {Peliti}}, \
  and\ \bibinfo {author} {\bibfnamefont {G.}~\bibnamefont {Parisi}},\
  }\bibfield  {title} {\enquote {\bibinfo {title} {Theoretical predictions for
  critical exponents at the $\lambda$-point of bose liquids},}\ }\href@noop {}
  {\bibfield  {journal} {\bibinfo  {journal} {Lett. Nuovo Cimento}\ }\textbf
  {\bibinfo {volume} {2}},\ \bibinfo {pages} {878--880} (\bibinfo {year}
  {1971})}\BibitemShut {NoStop}%
\bibitem [{\citenamefont {Charbonneau}\ and\ \citenamefont
  {Zamponi}(2022{\natexlab{b}})}]{parisi2022}%
  \BibitemOpen
  \bibfield  {author} {\bibinfo {author} {\bibfnamefont {P.}~\bibnamefont
  {Charbonneau}}\ and\ \bibinfo {author} {\bibfnamefont {F.}~\bibnamefont
  {Zamponi}},\ }\bibfield  {title} {\enquote {\bibinfo {title} {History of
  {RSB} interview: {G}iorgio {P}arisi},}\ }\href {\doibase
  10.34847/nkl.7fb7b5zw} {\bibfield  {journal} {\bibinfo  {journal} {History of
  RSB Project, CAPH\'ES, \'Ecole normale sup\'erieure, Paris}\ }\textbf
  {\bibinfo {volume} {80 p.}} (\bibinfo {year} {2022}{\natexlab{b}}),\
  10.34847/nkl.7fb7b5zw},\ \bibinfo {note} {{T}ranscript of an oral history
  conducted 2021 by Patrick Charbonneau and Francesco Zamponi}\BibitemShut
  {NoStop}%
\bibitem [{\citenamefont {Lubensky}\ and\ \citenamefont
  {Isaacson}(1978)}]{lubensky1978}%
  \BibitemOpen
  \bibfield  {author} {\bibinfo {author} {\bibfnamefont {T.~C.}\ \bibnamefont
  {Lubensky}}\ and\ \bibinfo {author} {\bibfnamefont {J.}~\bibnamefont
  {Isaacson}},\ }\bibfield  {title} {\enquote {\bibinfo {title} {Field theory
  for the statistics of branched polymers, gelation, and vulcanization},}\
  }\href {\doibase 10.1103/PhysRevLett.41.829} {\bibfield  {journal} {\bibinfo
  {journal} {Phys. Rev. Lett.}\ }\textbf {\bibinfo {volume} {41}},\ \bibinfo
  {pages} {829--832} (\bibinfo {year} {1978})}\BibitemShut {NoStop}%
\bibitem [{\citenamefont {Lubensky}(1975)}]{lubensky1975}%
  \BibitemOpen
  \bibfield  {author} {\bibinfo {author} {\bibfnamefont {T.~C.}\ \bibnamefont
  {Lubensky}},\ }\bibfield  {title} {\enquote {\bibinfo {title} {Critical
  properties of random-spin models from the $\ensuremath{\epsilon}$
  expansion},}\ }\href {\doibase 10.1103/PhysRevB.11.3573} {\bibfield
  {journal} {\bibinfo  {journal} {Phys. Rev. B}\ }\textbf {\bibinfo {volume}
  {11}},\ \bibinfo {pages} {3573--3580} (\bibinfo {year} {1975})}\BibitemShut
  {NoStop}%
\bibitem [{\citenamefont {Harris}\ \emph {et~al.}(1976)\citenamefont {Harris},
  \citenamefont {Lubensky},\ and\ \citenamefont {Chen}}]{lubensky1976}%
  \BibitemOpen
  \bibfield  {author} {\bibinfo {author} {\bibfnamefont {A.~B.}\ \bibnamefont
  {Harris}}, \bibinfo {author} {\bibfnamefont {T.~C.}\ \bibnamefont
  {Lubensky}}, \ and\ \bibinfo {author} {\bibfnamefont {J.-H.}\ \bibnamefont
  {Chen}},\ }\bibfield  {title} {\enquote {\bibinfo {title} {Critical
  properties of spin-glasses},}\ }\href {\doibase 10.1103/PhysRevLett.36.415}
  {\bibfield  {journal} {\bibinfo  {journal} {Phys. Rev. Lett.}\ }\textbf
  {\bibinfo {volume} {36}},\ \bibinfo {pages} {415--418} (\bibinfo {year}
  {1976})}\BibitemShut {NoStop}%
\bibitem [{\citenamefont {Harris}\ and\ \citenamefont
  {Lubensky}(1974)}]{harris1974}%
  \BibitemOpen
  \bibfield  {author} {\bibinfo {author} {\bibfnamefont {A.~B.}\ \bibnamefont
  {Harris}}\ and\ \bibinfo {author} {\bibfnamefont {T.~C.}\ \bibnamefont
  {Lubensky}},\ }\bibfield  {title} {\enquote {\bibinfo {title}
  {Renormalization-group approach to the critical behavior of random-spin
  models},}\ }\href {\doibase 10.1103/PhysRevLett.33.1540} {\bibfield
  {journal} {\bibinfo  {journal} {Phys. Rev. Lett.}\ }\textbf {\bibinfo
  {volume} {33}},\ \bibinfo {pages} {1540--1543} (\bibinfo {year}
  {1974})}\BibitemShut {NoStop}%
\bibitem [{\citenamefont {Parisi}(1979)}]{parisi1979}%
  \BibitemOpen
  \bibfield  {author} {\bibinfo {author} {\bibfnamefont {G.}~\bibnamefont
  {Parisi}},\ }\bibfield  {title} {\enquote {\bibinfo {title} {Toward a mean
  field theory for spin glasses},}\ }\href {\doibase
  10.1016/0375-9601(79)90708-4} {\bibfield  {journal} {\bibinfo  {journal}
  {Phys. Lett. A}\ }\textbf {\bibinfo {volume} {73}},\ \bibinfo {pages}
  {203--205} (\bibinfo {year} {1979})}\BibitemShut {NoStop}%
\bibitem [{\citenamefont {Young}(1979)}]{young1979}%
  \BibitemOpen
  \bibfield  {author} {\bibinfo {author} {\bibfnamefont {A.~P.}\ \bibnamefont
  {Young}},\ }\bibfield  {title} {\enquote {\bibinfo {title} {Fluctuation
  effects in spin glasses},}\ }\href {\doibase 10.1063/1.327239} {\bibfield
  {journal} {\bibinfo  {journal} {J. Appl. Phys.}\ }\textbf {\bibinfo {volume}
  {50}},\ \bibinfo {pages} {1691--1694} (\bibinfo {year} {1979})}\BibitemShut
  {NoStop}%
\bibitem [{\citenamefont {Charbonneau}(2021{\natexlab{k}})}]{young2021}%
  \BibitemOpen
  \bibfield  {author} {\bibinfo {author} {\bibfnamefont {P.}~\bibnamefont
  {Charbonneau}},\ }\bibfield  {title} {\enquote {\bibinfo {title} {History of
  {RSB} interview: {A.} {P}eter {Y}oung},}\ }\href {\doibase
  10.34847/nkl.2fef8760} {\bibfield  {journal} {\bibinfo  {journal} {History of
  RSB Project, CAPH\'ES, \'Ecole normale sup\'erieure, Paris}\ }\textbf
  {\bibinfo {volume} {20 p.}} (\bibinfo {year} {2021}{\natexlab{k}}),\
  10.34847/nkl.2fef8760},\ \bibinfo {note} {{T}ranscript of an oral history
  conducted 2021 by Patrick Charbonneau and Francesco Zamponi}\BibitemShut
  {NoStop}%
\bibitem [{\citenamefont {'t~Hooft}\ \emph {et~al.}(1980)\citenamefont
  {'t~Hooft}, \citenamefont {Itzykson}, \citenamefont {Jaffe}, \citenamefont
  {Lehmann}, \citenamefont {Mitter}, \citenamefont {Singer},\ and\
  \citenamefont {Stora}}]{cargese1980}%
  \BibitemOpen
  \bibinfo {editor} {\bibfnamefont {G.}~\bibnamefont {'t~Hooft}}, \bibinfo
  {editor} {\bibfnamefont {C.}~\bibnamefont {Itzykson}}, \bibinfo {editor}
  {\bibfnamefont {A.}~\bibnamefont {Jaffe}}, \bibinfo {editor} {\bibfnamefont
  {H.}~\bibnamefont {Lehmann}}, \bibinfo {editor} {\bibfnamefont {P.~K.}\
  \bibnamefont {Mitter}}, \bibinfo {editor} {\bibfnamefont {I.~M.}\
  \bibnamefont {Singer}}, \ and\ \bibinfo {editor} {\bibfnamefont
  {R.}~\bibnamefont {Stora}},\ eds.,\ \href@noop {} {\emph {\bibinfo {title}
  {Recent Developments in Gauge Theories: Proceedings of the NATO Advanced
  Study Institute of Recent Developments in Gauge Theories eld in Carg\`ese,
  Corsica, August 26-September 8, 1979}}}\ (\bibinfo  {publisher} {Plenum
  Press},\ \bibinfo {year} {1980})\BibitemShut {NoStop}%
\bibitem [{\citenamefont {Charbonneau}(2021{\natexlab{l}})}]{sourlas2021}%
  \BibitemOpen
  \bibfield  {author} {\bibinfo {author} {\bibfnamefont {P.}~\bibnamefont
  {Charbonneau}},\ }\bibfield  {title} {\enquote {\bibinfo {title} {History of
  {RSB} interview: {N}icolas {S}ourlas},}\ }\href {\doibase
  10.34847/nkl.2a55p6c3} {\bibfield  {journal} {\bibinfo  {journal} {History of
  RSB Project, CAPH\'ES, \'Ecole normale sup\'erieure, Paris}\ }\textbf
  {\bibinfo {volume} {23 p.}} (\bibinfo {year} {2021}{\natexlab{l}}),\
  10.34847/nkl.2a55p6c3},\ \bibinfo {note} {{T}ranscript of an oral history
  conducted 2021 by Patrick Charbonneau and Francesco Zamponi}\BibitemShut
  {NoStop}%
\bibitem [{\citenamefont {Parisi}(1980{\natexlab{b}})}]{parisi1980reports}%
  \BibitemOpen
  \bibfield  {author} {\bibinfo {author} {\bibfnamefont {G.}~\bibnamefont
  {Parisi}},\ }\bibfield  {title} {\enquote {\bibinfo {title} {Mean field
  theory for spin glasses},}\ }\href {\doibase 10.1016/0370-1573(80)90075-7}
  {\bibfield  {journal} {\bibinfo  {journal} {Phys. Rep.}\ }\textbf {\bibinfo
  {volume} {67}},\ \bibinfo {pages} {25--28} (\bibinfo {year}
  {1980}{\natexlab{b}})}\BibitemShut {NoStop}%
\bibitem [{\citenamefont {Thouless}\ \emph {et~al.}(1980)\citenamefont
  {Thouless}, \citenamefont {de~Almeida},\ and\ \citenamefont
  {Kosterlitz}}]{thouless1980}%
  \BibitemOpen
  \bibfield  {author} {\bibinfo {author} {\bibfnamefont {D.~J.}\ \bibnamefont
  {Thouless}}, \bibinfo {author} {\bibfnamefont {J.~R.~L.}\ \bibnamefont
  {de~Almeida}}, \ and\ \bibinfo {author} {\bibfnamefont {J.~M.}\ \bibnamefont
  {Kosterlitz}},\ }\bibfield  {title} {\enquote {\bibinfo {title} {Stability
  and susceptibility in {P}arisi's solution of a spin glass model},}\ }\href
  {\doibase 10.1088/0022-3719/13/17/017} {\bibfield  {journal} {\bibinfo
  {journal} {J. Phys. C}\ }\textbf {\bibinfo {volume} {13}},\ \bibinfo {pages}
  {3271--3280} (\bibinfo {year} {1980})}\BibitemShut {NoStop}%
\bibitem [{\citenamefont {M\'ezard}\ \emph {et~al.}(1987)\citenamefont
  {M\'ezard}, \citenamefont {Parisi},\ and\ \citenamefont
  {Virasoro}}]{mezard1987}%
  \BibitemOpen
  \bibfield  {author} {\bibinfo {author} {\bibfnamefont {M.}~\bibnamefont
  {M\'ezard}}, \bibinfo {author} {\bibfnamefont {G.}~\bibnamefont {Parisi}}, \
  and\ \bibinfo {author} {\bibfnamefont {M.A.}\ \bibnamefont {Virasoro}},\
  }\href@noop {} {\emph {\bibinfo {title} {Spin Glass Theory And Beyond: An
  Introduction To The Replica Method And Its Applications}}}\ (\bibinfo
  {publisher} {World Scientific Publishing Company},\ \bibinfo {year}
  {1987})\BibitemShut {NoStop}%
\bibitem [{\citenamefont {Anderson}(1989)}]{anderson1989}%
  \BibitemOpen
  \bibfield  {author} {\bibinfo {author} {\bibfnamefont {P.~W.}\ \bibnamefont
  {Anderson}},\ }\bibfield  {title} {\enquote {\bibinfo {title} {Spin glass
  {VI}: Spin glass as cornucopia},}\ }\href {\doibase 10.1063/1.2811137}
  {\bibfield  {journal} {\bibinfo  {journal} {Phys. Today}\ }\textbf {\bibinfo
  {volume} {42}},\ \bibinfo {pages} {9--11} (\bibinfo {year}
  {1989})}\BibitemShut {NoStop}%
\bibitem [{\citenamefont {Anderson}(1990)}]{anderson1990}%
  \BibitemOpen
  \bibfield  {author} {\bibinfo {author} {\bibfnamefont {P.~W.}\ \bibnamefont
  {Anderson}},\ }\bibfield  {title} {\enquote {\bibinfo {title} {Spin glass
  {VII}: Spin glass as paradigm},}\ }\href {\doibase 10.1063/1.2810479}
  {\bibfield  {journal} {\bibinfo  {journal} {Phys. Today}\ }\textbf {\bibinfo
  {volume} {43}},\ \bibinfo {pages} {9--11} (\bibinfo {year}
  {1990})}\BibitemShut {NoStop}%
\end{thebibliography}%

\end{document}